\documentclass[nofootinbib,superscriptaddress,a4paper,twocolumn,longbibliography,floatfix,pra]{revtex4-2}
\usepackage[english]{babel}
\usepackage[utf8]{inputenc}

\usepackage{amsthm}
\usepackage{mathtools}
\usepackage{physics}
\usepackage{xcolor}
\usepackage{graphicx}
\usepackage[left=23mm,right=13mm,top=35mm,columnsep=15pt]{geometry} 
\usepackage{adjustbox}
\usepackage{placeins}
\usepackage[T1]{fontenc}
\usepackage{lipsum}
\usepackage{csquotes}

\usepackage{listings}
\usepackage{xcolor}

\definecolor{codegreen}{rgb}{0,0.6,0}
\definecolor{codegray}{rgb}{0.5,0.5,0.5}
\definecolor{codepurple}{rgb}{0.58,0,0.82}
\definecolor{tqblue}{HTML}{08293d}
\definecolor{backcolour}{HTML}{fefdf5}

\lstdefinestyle{mystyle}{
    backgroundcolor=\color{backcolour},   
    commentstyle=\color{codegreen},
    keywordstyle=\color{magenta},
    numberstyle=\tiny\color{codegray},
    stringstyle=\color{codepurple},
    basicstyle=\ttfamily\footnotesize\color{tqblue},
    breakatwhitespace=false,         
    breaklines=true,
    postbreak=\mbox{\textcolor{magenta}{$\hookrightarrow$}\space},                 
    captionpos=b,                    
    keepspaces=true,                 
    numbers=left,                    
    numbersep=5pt,                  
    showspaces=false,                
    showstringspaces=false,
    showtabs=false,                  
    tabsize=2
}

\usepackage{booktabs}
\usepackage{soul}

\usepackage{xspace}
\newcommand{\tequila}{\textsc{tequila}\xspace}

\usepackage{xcolor}
\usepackage[pdftex, pdftitle={Article}, pdfauthor={Author},colorlinks=true,citebordercolor={1 0 0},linkbordercolor={0 0 1}]{hyperref} % For hyperlinks in the PDF
\hypersetup{
    colorlinks,
    linkcolor={red!50!black},
    citecolor={blue!50!black},
    urlcolor={blue!80!black}
}

\begin{document}
\title{\tequila: A platform for rapid development of quantum algorithms.}

\author{Jakob S. Kottmann}
\email[E-mail: ]{jakob.kottmann@utoronto.ca}
\affiliation{Chemical Physics Theory Group, Department of Chemistry, University of Toronto, Canada.}
\affiliation{Department of Computer Science, University of Toronto, Canada.}

\author{Sumner Alperin-Lea}
\email[E-mail: ]{sumner.alperin@mail.utoronto.ca}
\affiliation{Chemical Physics Theory Group, Department of Chemistry, University of Toronto, Canada.}

\author{Teresa Tamayo-Mendoza}
\affiliation{Department of Chemistry and Chemical Biology, Harvard University}
\affiliation{Chemical Physics Theory Group, Department of Chemistry, University of Toronto, Canada.}
\affiliation{Department of Computer Science, University of Toronto, Canada.}

\author{Alba Cervera-Lierta}
\affiliation{Chemical Physics Theory Group, Department of Chemistry, University of Toronto, Canada.}
\affiliation{Department of Computer Science, University of Toronto, Canada.}

\author{Cyrille Lavigne}
\affiliation{Chemical Physics Theory Group, Department of Chemistry, University of Toronto, Canada.}
\affiliation{Department of Computer Science, University of Toronto, Canada.}

\author{Tzu-Ching Yen}
\affiliation{Chemical Physics Theory Group, Department of Chemistry, University of Toronto, Canada.}

\author{Vladyslav Verteletskyi}
\affiliation{Chemical Physics Theory Group, Department of Chemistry, University of Toronto, Canada.}

\author{Philipp Schleich}
\affiliation{Center for Computational Engineering Science, RWTH Aachen University, Aachen, Germany.}

\author{Abhinav Anand}
\affiliation{Chemical Physics Theory Group, Department of Chemistry, University of Toronto, Canada.}

\author{Matthias Degroote}
\affiliation{Current address: Boehringer Ingelheim, Amsterdam, Netherlands}
\affiliation{Chemical Physics Theory Group, Department of Chemistry, University of Toronto, Canada.}
\affiliation{Department of Computer Science, University of Toronto, Canada.}

\author{Skylar Chaney}
\affiliation{Chemical Physics Theory Group, Department of Chemistry, University of Toronto, Canada.}
\affiliation{Department of Physics, Williams College, Williamstown, MA, USA.}

\author{Maha Kesibi}
\affiliation{Chemical Physics Theory Group, Department of Chemistry, University of Toronto, Canada.}
\affiliation{Department of Computer Science, University of Toronto, Canada.}

\author{Naomi Grace Curnow}
\affiliation{University of Cambridge, United Kingdom}

\author{Brandon Solo}
\affiliation{Quantum Open Source Foundation QC Mentorship Program}

\author{Georgios Tsilimigkounakis}
\affiliation{Quantum Open Source Foundation QC Mentorship Program}

\author{Claudia Zendejas-Morales}
\affiliation{Quantum Open Source Foundation QC Mentorship Program}

\author{Artur F. Izmaylov}
\affiliation{Chemical Physics Theory Group, Department of Chemistry, University of Toronto, Canada.}
\affiliation{Department of Physical and Environmental Sciences, University of Toronto Scarborough, Canada.}

\author{Alán Aspuru-Guzik}
\email[E-mail: ]{aspuru@utoronto.ca}
\affiliation{Chemical Physics Theory Group, Department of Chemistry, University of Toronto, Canada.}
\affiliation{Department of Computer Science, University of Toronto, Canada.}
\affiliation{Vector Institute for Artificial Intelligence, Toronto, Canada.}
\affiliation{Canadian  Institute  for  Advanced  Research  (CIFAR)  Lebovic  Fellow,  Toronto,  Canada.}

\date{\today} % Leave empty to omit a date

\begin{abstract}
   Variational quantum algorithms are currently the most promising class of algorithms for deployment on near-term quantum computers.  In contrast to classical algorithms, there are almost no standardized methods in quantum algorithmic development yet, and the field continues to evolve rapidly. As in classical computing, heuristics play a crucial role in the development of new quantum algorithms, resulting in a high demand for flexible and reliable ways to implement, test, and share new ideas. Inspired by this demand, we introduce \tequila, a development package for quantum algorithms in \textsc{python}, designed for fast and flexible implementation, prototyping and deployment of novel quantum algorithms in electronic structure and other fields.  \tequila operates with abstract expectation values which can be combined, transformed, differentiated, and optimized.  On evaluation, the abstract data structures are compiled to run on state of the art quantum simulators or interfaces. 
\end{abstract}

\maketitle

\section{Introduction}
Quantum computing is currently in the \textit{Noisy Intermediate-Scale Quantum} (NISQ)~\cite{preskill2018nisq,nisq_review} era, in which devices composed of a few qubits can execute non-trivial quantum computation while grappling with non-negligible noise and a lack of error correction. 
The so-called hybrid quantum-classical algorithms, and, in particular, the variational quantum algorithms (VQAs), constitute one of the most relevant NISQ algorithms subset. Such algorithms leverage classical coprocessors to iteratively improve the performance of parametrized quantum circuits with respect to a variety of objectives. VQAs exhibit generally shallow depth and are postulated to have greater resistance to noise, both of which are necessary to extract the potential of near term devices. Algorithms like the Variational Quantum Eigensolver (VQE)~\cite{peruzzo2014variational} and the Quantum Approximate Optimization Algorithm (QAOA)~\cite{farhi2014qaoa} have shown great promise in solving difficult problems in a variety of fields, including quantum chemistry, materials science, finance and quantum machine learning~\cite{wittek2014qml}. 

The rapid expansion of VQAs and QML algorithms has been accompanied and assisted by an ever-expanding market of quantum information and simulation packages. To date, over 100 different quantum simulation packages have been released.~\cite{quantiki}. Many of the \textit{full stack} simulation packages are written in \textsc{python}, and some of them are backed by leading companies in the quantum industry, such as \textsc{cirq}~\cite{cirq}, \textsc{qiskit}~\cite{Qiskit},\textsc{q\#}~\cite{q_sharp}, \textsc{pyquil}~\cite{smith2016practical} and \textsc{Strawberry Fields}~\cite{Killoran2019strawberryfields} respectively developed by or affiliated with Google, IBM, Microsoft, Rigetti, and Xanadu. Where applicable, these industrially developed packages also integrate the cloud quantum computing services made available by their respective organization. 

However, the large variety of simulation software poses a challenge both for the validation of experiments and the adoption of new algorithms. As the NISQ era continues to progress, quantum scientists stand to benefit markedly from a unified development framework in which the strengths and resources of different software packages -- both quantum and classical -- can be easily coordinated, with minimal constriction of algorithmic design choices, to further accelerate the pace of development. To this end, we introduce \tequila.

We note that a number of software packages have been created to meet similar needs, such as \textsc{Pennylane}~\cite{bergholm2018pennylane}, which made a pioneering step in introducing automatic differentiation to variational quantum algorithms, as well as classical quantum simulators with an extended interface, including \textsc{Xacc}~\cite{alex2019xacc}, \textsc{ProjectQ}~\cite{Steiger2018projectqopensource}, \textsc{Qibo}~\cite{efthymiou2020qibo},
\textsc{Yao}~\cite{YaoFramework2019},
\textsc{QuaSiMo}/\textsc{Qcor}~\cite{nguyen2021composable},
and \textsc{Qulacs}~\cite{qulacs}, alongside several previously mentioned packages  (see a recent review~\cite{nisq_review} for an overview). \tequila differs from the aforementioned packages either by a difference in functionality or through its application programming interface (API). The benefits and strengths of individual packages largely depend on the intended applications and individual preferences of the user.

\tequila is an open-source \textsc{python} 3 software package, which integrates diverse simulation software, classical optimization routines, and powerful tools for the manipulation and combination of quantum circuits and variational objectives thereof. Additionally, \tequila has a native interface for electronic structure packages like \textsc{psi4}~\cite{psi4}, \textsc{pyscf}~\cite{sun2018pyscf,sun_secondary_pyscf} and \textsc{madness}~\cite{harrison2016madness}. Focused on variational algorithms, whose objectives may require classical transformations on the output of expectation values or circuit measurements, \tequila implements convenient tools for arithmetical operations on those structures, and interfaces the powerful autodifferentiation libraries \textsc{jax}~\cite{jax} and \textsc{autograd} to allow hassle-free analytic differentiation of user-defined objective functions. \tequila maintains a firmly object-oriented user interface, in which circuits, Hamiltonians, expectation values, and user-defined objectives can be conveniently combined arithmetically, using code that represents the underlying mathematics in a "blackboard" fashion. 

In this article, we detail examples of how construction, compilation, manipulation, differentiation, and optimization of variational quantum objectives can be performed in \tequila to illustrate the core of the application programming interface. Further tutorials and documentation are available on the \tequila \href{https://github.com/aspuru-guzik-group/tequila}{Github repository}~\cite{tequila}. We first present {\tequila}s API and how it deals with the abstract data structures in the form of objectives, Hamiltonians and quantum circuits, and the simulation and execution of these structures. We then proceed to show particular usage examples that include quantum chemistry simulation and a quantum machine learning application. Finally, we close this paper with some conclusions and the further development of this quantum software package.

\section{How \tequila works}

\begin{figure}
    \centering
    \begin{tabular}{c}
    \includegraphics[width=0.45\textwidth]{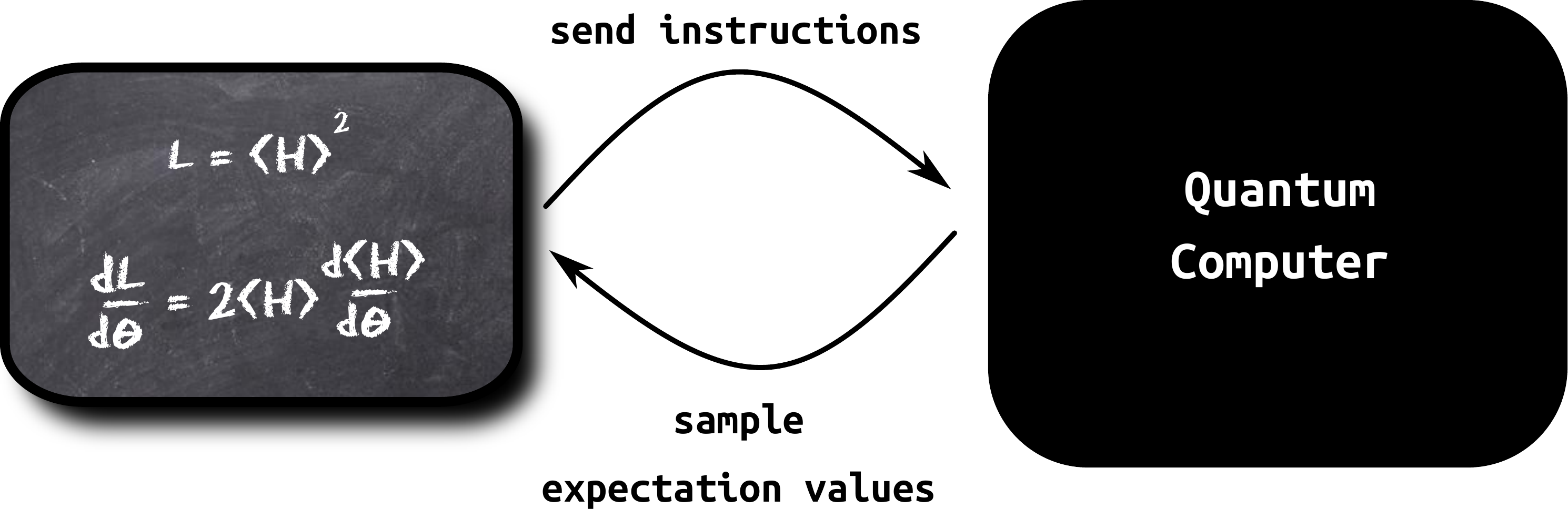}\\
    \\
    \midrule
    \\
    \includegraphics[width=0.45\textwidth]{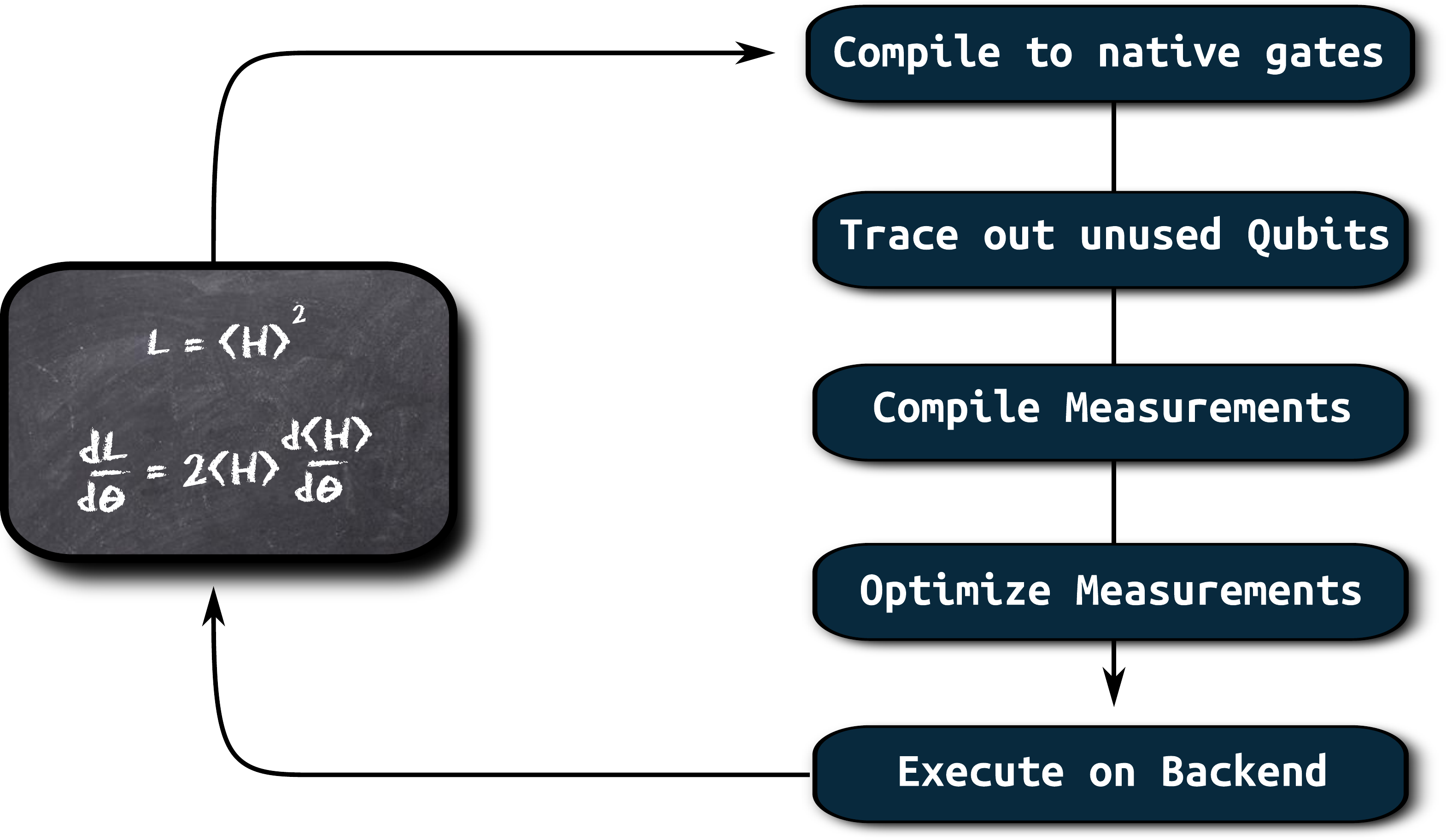}
    \end{tabular}
    \caption{Abstract illustration of black-box and explicit usage of quantum computer simulation or execution backends in \tequila. (Top) Abstract algorithm development, treating the quantum backend as a black-box sampling device for expectation values. (Bottom) Illustration of the underlying modularized machinery that handles communication with the quantum backend.}
    \label{fig:black_box_qc}
\end{figure}

The core intention of \tequila, is to provide an open-source environment for the rapid development and demonstration of new ideas in quantum computation, with a focus on variational algorithms. \tequila is designed through a high level of abstraction, resembling the chalk-and-blackboard mathematics underlying the algorithm and inspired by the API of the \textsc{madness}~\cite{harrison2016madness} package.
The user is offered a choice, illustrated in Fig.~\ref{fig:black_box_qc}, between treating quantum computers as black box samplers of abstract expectation values, or controlling the use of the quantum computer more directly.
In the following, we will describe how \tequila may be used in both fashions.

\subsection{Abstract data structures}

\begin{figure}
    \centering
    \includegraphics[width=0.45\textwidth]{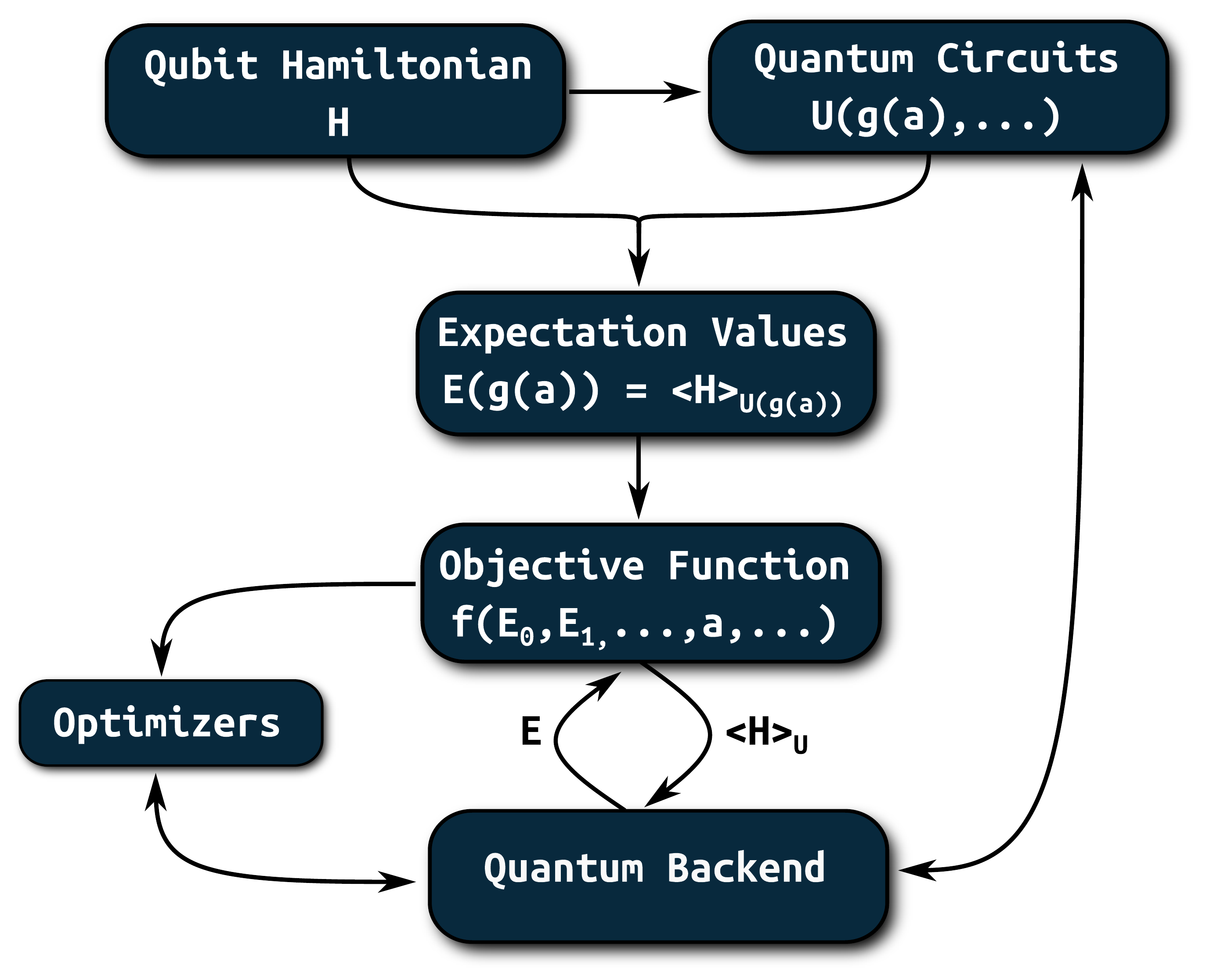}
    \caption{Abstract data structures in \tequila. Qubit Hamiltonians (hermitian operators) define measurements in an expectation value, or generate a (parametrized) unitary operation within a quantum circuit. Objectives can be formed by arbitrarily transformed expectation values which can be automatically evaluated through several supported quantum backends.}
    \label{fig:data_structures}
\end{figure}

\tequila allows users to treat quantum backends, which serve as interfaces to real hardware or simulators, as abstract sampling devices, requiring little to no knowledge about the underlying technical details of simulation or execution. The user merely needs to be familiar with the physical principles of quantum computation, and have an idea about how the specific problems of interest could be solved with access to quantum computers.
The core functionality of \tequila is provided by abstract data structures depicted in Fig.~\ref{fig:data_structures} where the user only deals directly with \texttt{Objectives}, \texttt{Hamiltonians} and \texttt{Circuits}.
These last two objects define abstract Hamiltonians and unitaries (quantum circuits) which can be combined to create abstract expectation values.
\tequila bundles these abstract expectation values into objectives which can then be transformed, combined or differentiated in a blackboard-style fashion.

\subsubsection{Objectives}\label{sec:objectives}

Objectives are callable data structures that hold a list containing abstract expectation values and variables alongside a transformation that defines how those structures shall be processed after evaluation.
Formally, a \tequila objective can be written as
\begin{align}
    O = f\left(E_0, E_1, \dots, a_0, a_1, \dots\right),
\end{align}
with variables $a_k$ and expectation values $E_k$. These expectation values also depend on the variables $a_{k}$ through the unitary circuits $U_{k}$,
\begin{align}
E_k &= \bra{0}U_k^\dagger\left(g(a,b,\dots)\right) H_k U_k\left(g(a,b,\dots)\right)\ket{0}
\end{align}
where the variables can be potentially transformed by arbitrary transformations $g$.

\begin{figure}
    \begin{tabular}{c}
    \includegraphics[width=0.5\textwidth]{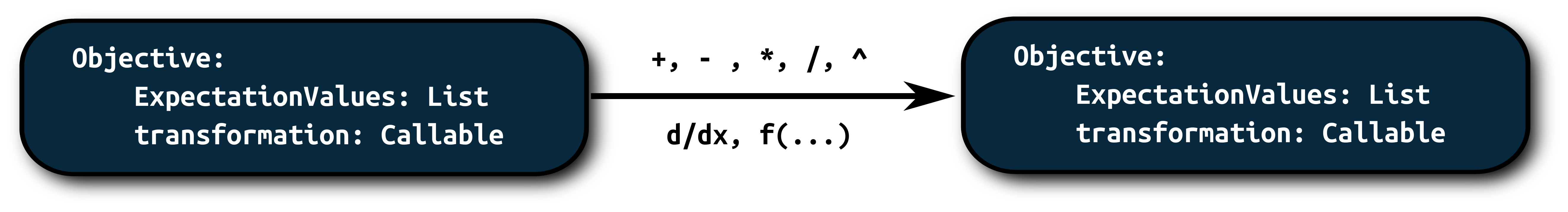}
    \\
    \midrule
    \\
    \end{tabular}
\begin{lstlisting}[language=Python]
# initialize expectation values
E0 = tq.ExpectationValue(H=H0, U=U0)
E1 = tq.ExpectationValue(H=H1, U=U1)
E2 = tq.ExpectationValue(H=H2, U=U2)

# combine them
O1 = E0 + E1
O2 = 0.5*E2**2
O3 = O1**O2
\end{lstlisting}
    \begin{tabular}{c}
    \includegraphics[width=0.5\textwidth]{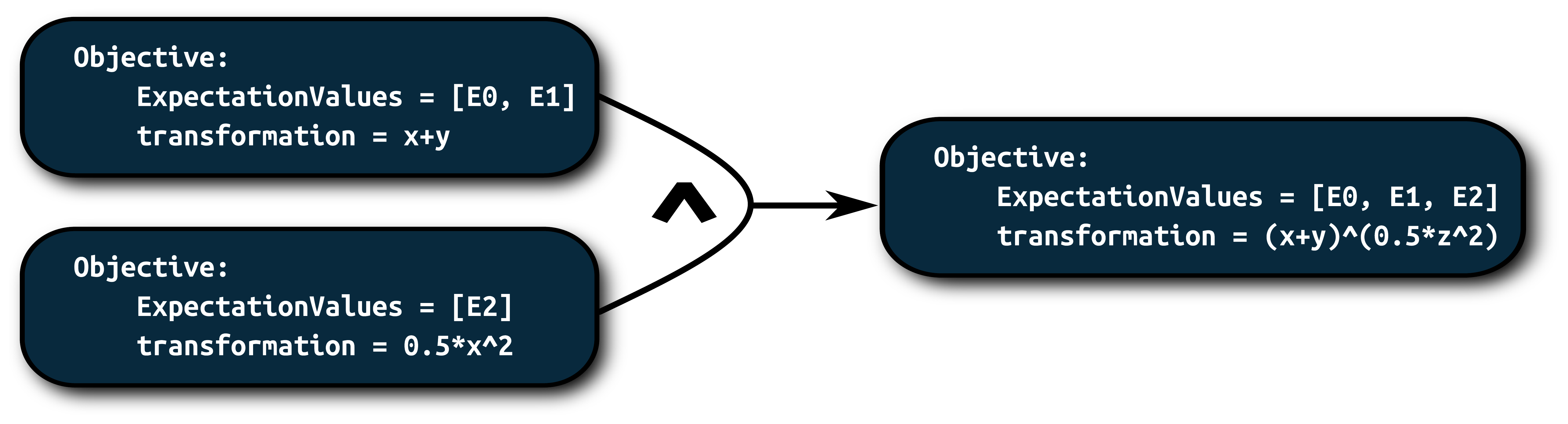}
    \end{tabular}
    \caption{Top: Abstract data structure for objectives in \tequila. Arithmetic operations, transformations or derivatives give back objects of the same type. Bottom: Example of arithmetic operations on \tequila objectives and their resulting data structures. The right hand side shows the full objective defined as $O_3 = (E_0 + E_1)^{\frac{1}{2}E_2^2}$ }
    \label{fig:data_structures_objectives_arithmetics}
\end{figure}

Operations on objectives return objects of the same type, as illustrated in Fig.~\ref{fig:data_structures_objectives_arithmetics}, rendering the combination and extension of objectives more convenient.
As example, consider the addition of the two objectives $O_1 = E_1^2 $ and the more complicated $O_2 = e^{-E_2^2} + c$. In this example, each of these objectives only carries one expectation value for simplicity and $c$ can be a constant, a variable or a scalar function. The corresponding transformations in the data structures are $f_1(x) = x^2 $ and $f_2(x) = e^{-x^2} + c$. If both objectives are added together as $O_3 = O_1 + O_2$, the resulting data structure carries two abstract expectation values, and the transformation becomes $f_3(x,y) = x^2 + e^{-y^2} + c$. See Fig.~\ref{fig:data_structures_objectives_arithmetics} for an explicit illustration of a similar example.

\begin{figure}
    \centering
    \begin{tabular}{l}
    \includegraphics[width=0.45\textwidth]{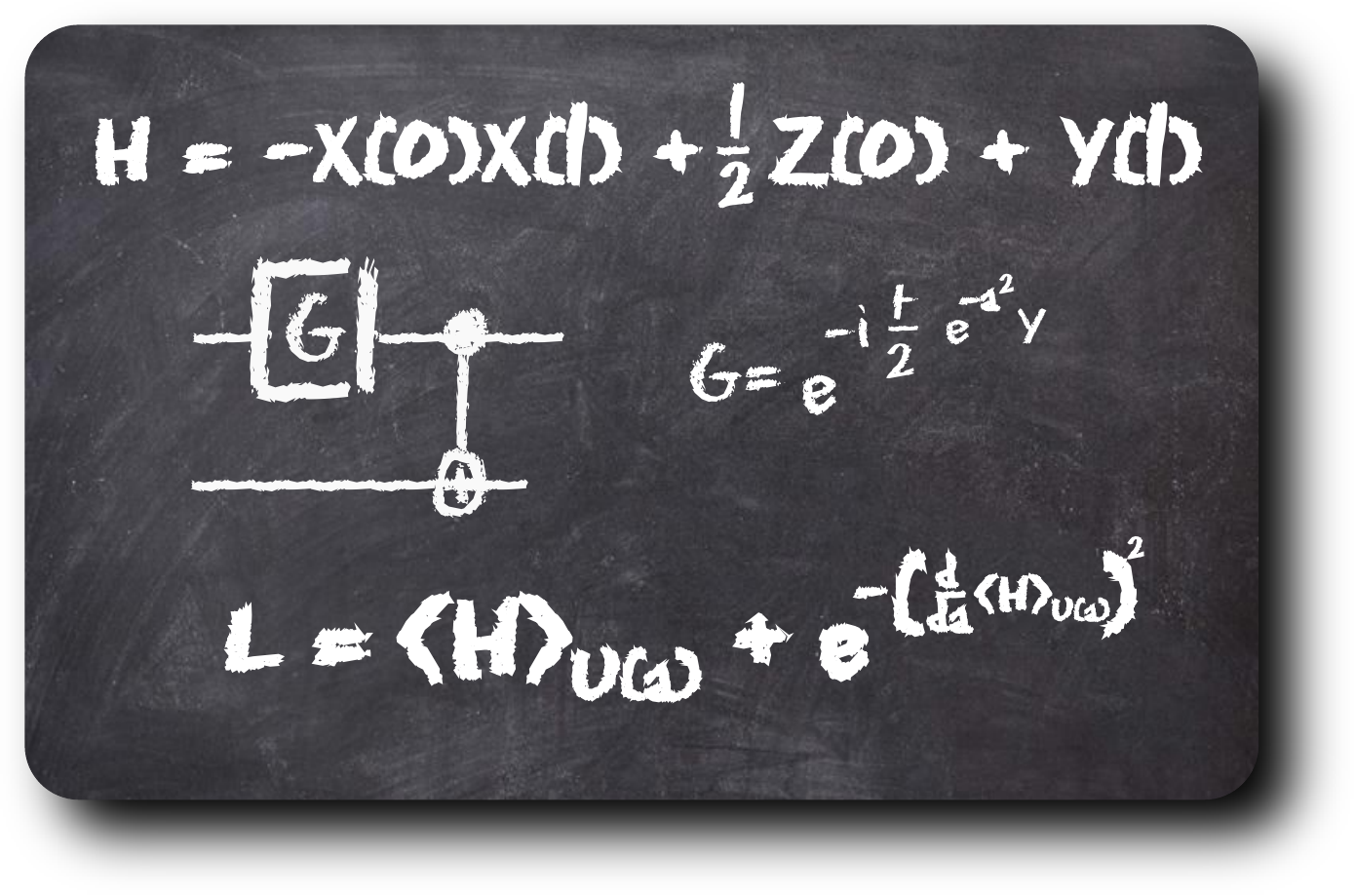}\\
\begin{lstlisting}[language=Python]
import tequila as tq
from numpy import exp, pi
a = tq.Variable("a")
U = tq.gates.Ry(angle=(-a**2).apply(exp)*pi, target=0)
U += tq.gates.X(target=1, control=0)
H = tq.QubitHamiltonian.from_string("-1.0*X(0)X(1)+0.5Z(0)+Y(1)")
E = tq.ExpectationValue(H=H, U=U)
dE = tq.grad(E, "a")
objective = E + (-dE**2).apply(exp)
result = tq.minimize(method="phoenics", objective=objective)
\end{lstlisting}\\
    \includegraphics[width=0.45\textwidth]{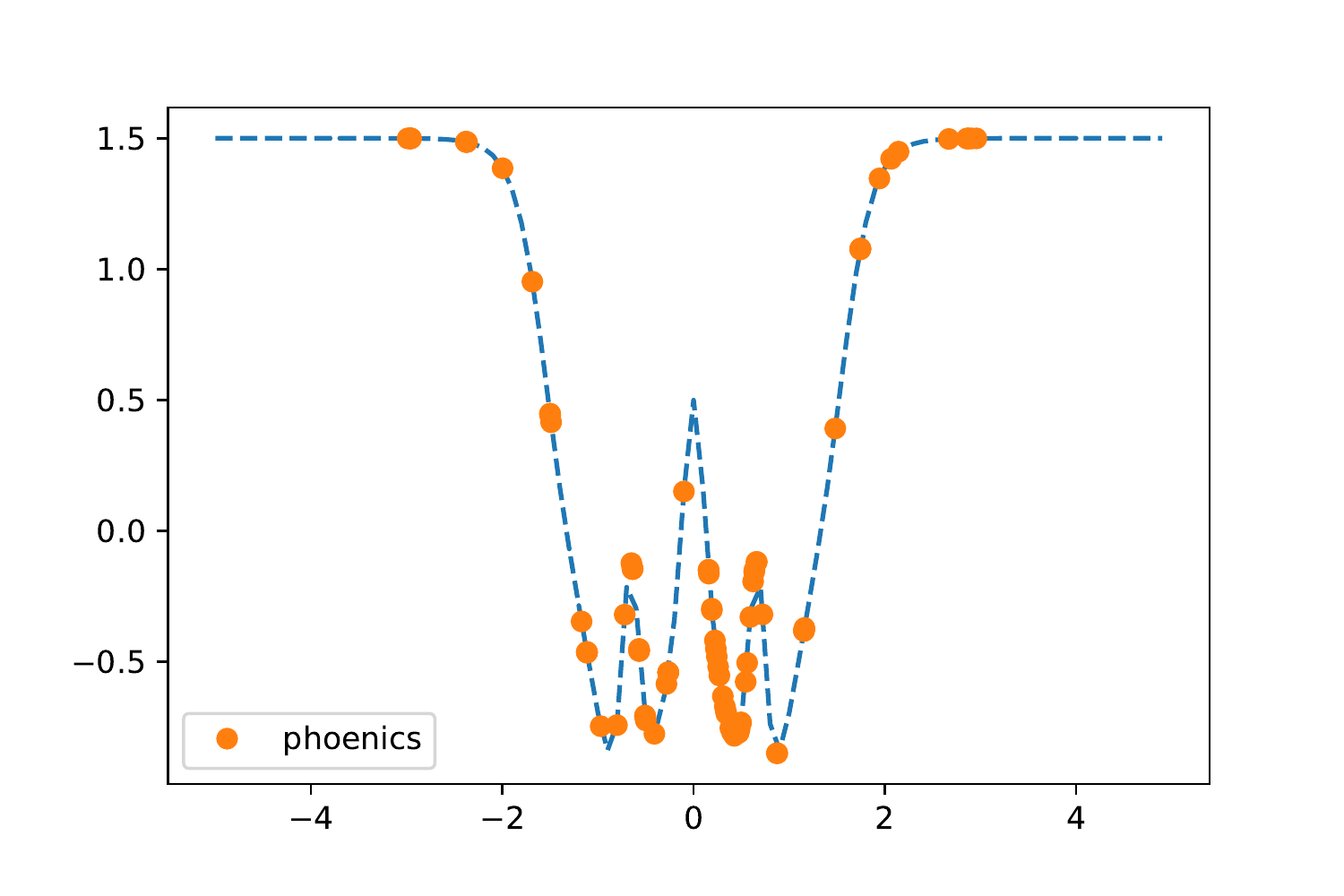}
    \end{tabular}
    \caption{An illustrative toy model implemented and optimized within \tequila. The top panel illustrates the model system used, where  defined loss function $L$ was minimized. The middle panel depicts the implementation with \tequila. The lower panel depicts an example of optimization results using the \textsc{phoenics} optimizer.}
    \label{fig:example_all_in_one}
\end{figure}

\begin{figure}
    \centering
\begin{lstlisting}[language=Python]
U = tq.gates.H(0)
U += tq.gates.Ry("a", target=1, control=0)
O = tq.ExpectationValue(U=U, H=H)

dO = tq.grad(O, "a")
\end{lstlisting}
    \begin{tabular}{c}
    \toprule
    \\
    \includegraphics[scale=0.8]{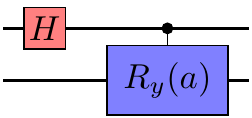}\\
    \includegraphics[scale=0.8]{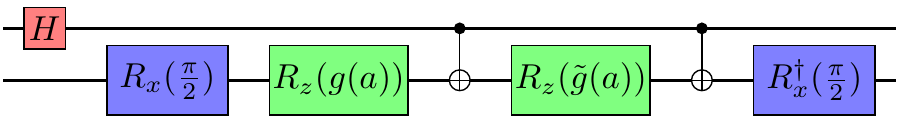}\\
    \\
    \midrule
    \\
    \includegraphics[width=0.45\textwidth]{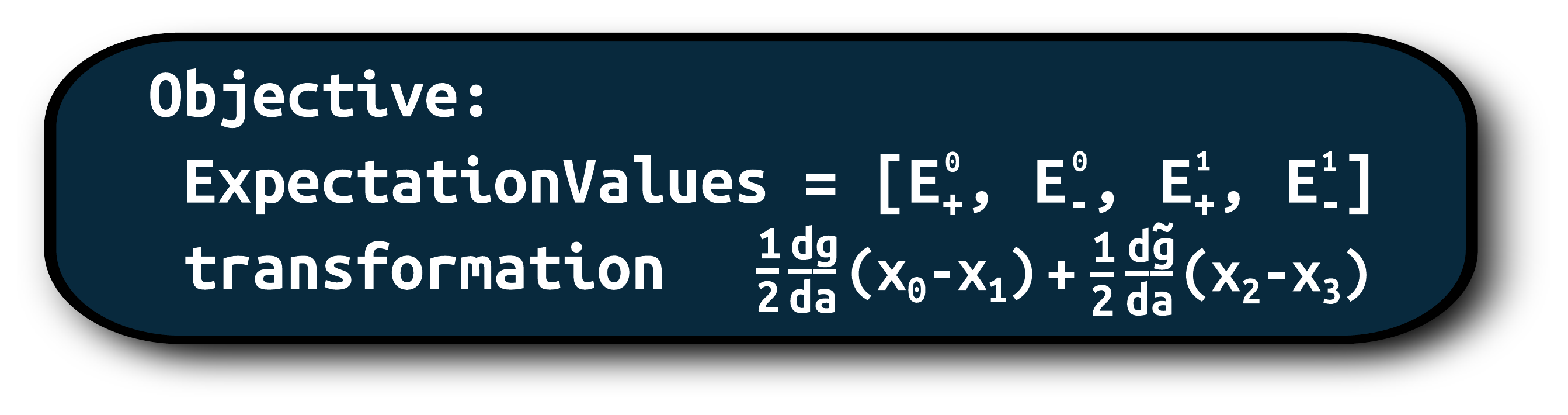}
    \end{tabular}
    \caption{Gradient compiling in \tequila: (Top) \tequila Top level code. (Middle) The abstract circuit and the internally compiled circuit.
    (Bottom) The objective structure representing the gradient $dO$ in the code. $E^i_\pm$ denote the left and right shifted expectation values with respect to the $i$th occurence of the variable $a$ in the compiled circuit. In this specific example the transformations $g$ and $\tilde{g}$ are rescaling by $\pm\frac{1}{2}$.}
    \label{fig:data_structures_objectives_gradients}
\end{figure}

\tequila objectives are fully differentiable. We make use of the \textit{shift-rule} technique developed by Schuld \textit{et.~al.}~\cite{schuld2019evaluating}, first implemented within \textsc{pennylane}~\cite{bergholm2018pennylane}, alongside automatic differentiation of the transformation of the objective. For objectives whose expectation values contain parametrized quantum gates that do not fulfill the requirements of being directly differentiable, we employ a variety of decomposition techniques, similar in spirit to those proposed in Ref.~\cite{crooks2019gradients}. Differentiation of an \texttt{Objective} again gives back an \texttt{Objective}, enabling convenient access to arbitrary order derivatives. An explicit illustration using automatic decomposition of controlled rotations and the resulting \texttt{Objective} data structure is shown in Figs.~\ref{fig:data_structures_objectives_gradients} and~\ref{fig:example_all_in_one} , where we illustrate blackboard-style \textsc{tequila} code for a small toy model.

\tequila objectives are abstract data structures that can be translated into various quantum backends which are interfaces to existing quantum hardware or simulators. The \texttt{compile} function allows the translation of an abstract \texttt{Objective} into an \texttt{Objective} that is tied to a specific backend. After compilation, the \texttt{objective} can be used as an abstract function with respect to its parameters. In the following code snipped, we present a small example of how the second derivative of an \texttt{Objective} can be obtained as an abstract \texttt{objective}, which can then be translated to a quantum backend and later on be used as an abstract function:

\begin{lstlisting}[language=Python]
import tequila as tq
import numpy as np

# a trivial example
objective=tq.Variable("a").apply(np.cos) 

# first derivative
dO = tq.grad(objective, "a")
# second derivative
ddO = tq.grad(dO, "a")

# compiled second derivative
compiled = tq.compile(ddO)

# use like a function
evaluated = compiled({"a": 1.0})
\end{lstlisting}

By not setting the \texttt{samples} keyword in the evaluation, the result will become the exact simulation of all expectation values. The original objective could be, for example, the objective in Fig.~\ref{fig:example_all_in_one}.
Note that the quantum backend was not specified in the above code example. If this is the case, \tequila will detect all supported and installed backends automatically and choose the one most appropriate for the given task based on an intrinsically defined hierarchy of efficiency for that task.
The \texttt{compile} function takes additional keywords like \texttt{backend}, \texttt{samples}, and \texttt{noise} in order to specify which backend to use, if finite samples are simulated or if a noise model is used for a simulation. If finite samples are desired, those can be passed to the \texttt{compile} function, influencing the automatic selection of the available quantum backends. After compilation, the sample size can still be changed when calling the \texttt{Objective}.

\subsubsection{Hamiltonians}\label{sec:hamiltonians}

Hamiltonians in \tequila are represented as linear combinations of tensor products of Pauli matrices -- so called Pauli strings -- and by default use \textsc{openfermion}\cite{openfermion} as a backend for algebraic manipulations.
Transformations into different data formats like the \textit{symplectic binary form} are also possible for more in-depth tasks (see the next section). In principle, every Hamiltonian class that is able to provide a list of abstract Pauli strings can be used within the \tequila architecture.

Within the abstract expectation values, the Hamiltonians represent averaged measurements. 
A measurement of a single qubit in the computational basis can, for example, be represented by the two projectors
\begin{align}
    Q_+ &= \ket{0}\bra{0} = \frac{1}{2}\left(1 + \sigma_z\right)\label{eq:Qp} \\
    Q_- &= \ket{1}\bra{1} = \frac{1}{2}\left(1 - \sigma_z\right)\label{eq:Qm}.
\end{align}

Hamiltonians can be initialized directly, through strings, 
%through explicitly written matrices, % code not integrated
or from \textsc{openfermion} operators.
Take for example the following Hamiltonian acting on three qubits (specified in parenthesis),
\begin{align}
    H = \sigma_x(0) \sigma_y(1) + 3\sigma_y(3).
\end{align}
It can be initialized by combining primitives as
\begin{lstlisting}[language=Python]
H = tq.paulis.X(0)*tq.paulis.Y(1)
H += 3.0*tq.paulis.Y(3)
\end{lstlisting}
or from strings,
\begin{lstlisting}[language=Python]
H = tq.QubitHamiltonian.from_string("1.0*X(0)*Y(1) + 3.0*Y(3)")
\end{lstlisting}

More sophisticated Hamiltonians, such as that of the Heisenberg model,
\begin{align}
H &= \sum_k J_x \sigma_x(k)\sigma_x(k+1)\nonumber \\ 
&+\sum_k J_y\sigma_y(k)\sigma_y(k+1) \nonumber \\ &+\sum_k J_z\sigma_z(k)\sigma_z(k+1)\nonumber \\ 
&+\sum_k h \sigma_z(k),
\end{align}
can be initialized in a similar way:
\begin{lstlisting}[language=Python]
H = tq.paulis.Zero()
for k in range(n_qubits):
    H += Jx*tq.paulis.X([k,k+1])
    H += Jx*tq.paulis.Y([k,k+1])
    H += Jx*tq.paulis.Z([k,k+1])
    H += h*tq.paulis.Z([k])
\end{lstlisting}
where \texttt{Jx,Jy,Jz} and \texttt{h} are floating point numbers and the integer \texttt{n\_qubits} determines the number of sites.

Hamiltonians can also be defined more indirectly by defining abstract qubit wave functions and forming operations on them in Dirac notation.
Take for example the projector on a predefined wave function
\begin{align}
    H = \ket{\Psi}\bra{\Psi}
\end{align}
which can be defined within \tequila as
\begin{lstlisting}[language=Python]
wfn = tq.QubitWaveFunction("1.0|00> + 1.0|11>")
wfn = wfn.normalize()
H =  tq.paulis.Projector(wfn)
\end{lstlisting}
where \texttt{wfn} is the $\ket{\Psi}$ wave function that can be initialized from strings or arrays of coefficients $c_i$ that corresponds to the computational basis state $\ket{i}$ in binary notation. The internal representation of the projector after initialization is decomposed into Pauli strings so that we end up with the same data structure as before. 
As example, consider the Bell state wavefunction
\begin{align}
    \ket{\Psi^+} = \frac{1}{\sqrt{2}}\left( \ket{00} + \ket{11} \right),
\end{align}
from the code snipped above. The projector takes the form
\begin{align}
    \ket{\Psi^+}\bra{\Psi^+} =& \frac{1}{2}\left(\vphantom{\frac{1}{2}} \ket{00}\bra{00} + \ket{11}\bra{00} \right.\nonumber\\
    &+ \left.\vphantom{\frac{1}{2}} \ket{00}\bra{11} + \ket{11}\bra{11} \right),
\end{align}
and each of the four individual terms is decomposed into Pauli strings, similarly to Ref.~\cite{nicolas2019}, by using Eqs.~\eqref{eq:Qp} and ~\eqref{eq:Qm} and
\begin{align}
    \sigma^+ &\equiv \ket{0}\bra{1} = \frac{1}{2}\left( \sigma_x + i\sigma_y \right), \\
    \sigma^- &\equiv \ket{1}\bra{0} = \frac{1}{2}\left( \sigma_x - i\sigma_y \right). 
\end{align}

Similar to \texttt{Projector}, the \texttt{KetBra} function of \tequila decomposes the operator $\ket{\Psi}\bra{\Phi}$ into a non-hermitian combination of Pauli strings. An arbitrary matrix can then be decomposed into a \tequila Hamiltonian with the help of the \texttt{KetBra} function. By applying the \texttt{split} function, the anti-hermitian and hermitian parts can be separated for individual treatment within abstract expectation values, as shown in this code example for a given $N\times M$ \texttt{matrix}
\begin{lstlisting}[language=Python]
n_qubits = np.ceil(np.log2(max(matrix.shape)))
n_qubits = int(n_qubits)

H = tq.paulis.Zero()
for i in range(matrix.shape[0]):
    for j in range(matrix.shape[1]):
        H += matrix[i,j]*tq.paulis.KetBra(ket=i, bra=j, n_qubits=n_qubits)

hermitian, anti = H.split()
\end{lstlisting}
where the matrix is encoded in binary into $\lceil{\log_2\left(\max\left(M,N\right)\right)}\rceil$ qubits.

\begin{figure}
    \centering
    \includegraphics[width=0.4\textwidth]{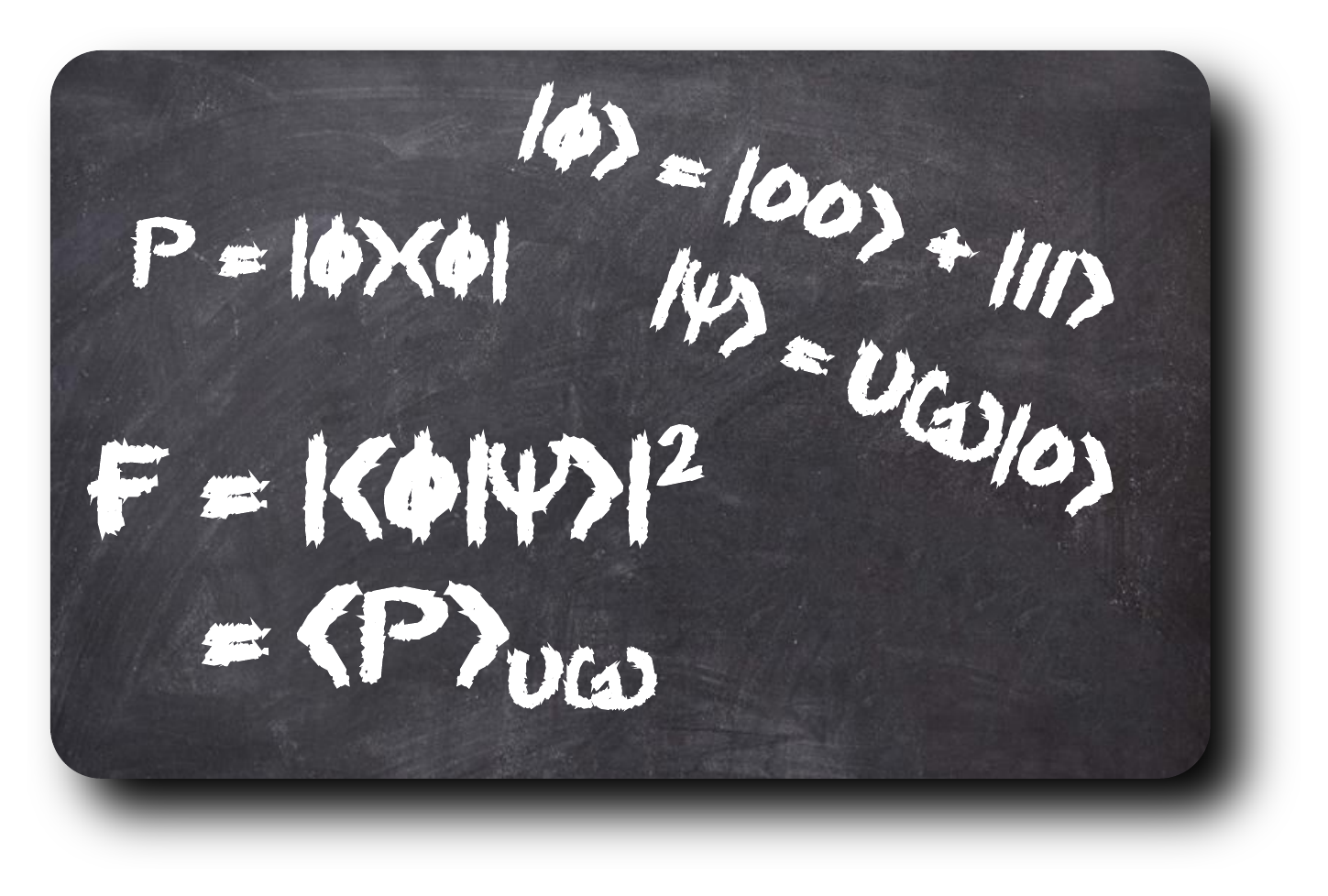}
\begin{lstlisting}[language=Python]
# Fidelity with a known state
# Bell state
wfn = tq.QubitWaveFunction.from_string(
      "1.0*|00> + 1.0*|11>")
wfn = wfn.normalize()

# compile as objective
P = tq.paulis.Projector(wfn)
O = tq.ExpectationValue(H=P, U=U)
F = tq.simulate(O)

# Fidelity with the Bell state
# encoded in circuit U2
U2 = tq.gates.H(0)
U2 += tq.gates.CNOT(0,1)
P0 = tq.paulis.Unit()
for n in range(U2.n_qubits):
    P *= tq.paulis.Qp(n)
O = tq.ExpectationValue(H=P0, U=U+U2.dagger())
F = tq.simulate(O)
\end{lstlisting}        
    \caption{Example portraying how to obtain the fidelity between two states using \tequila and different strategies. One for states that are analytically known and one for states encoded in quantum circuits. We use the same Bell state for both illustrations. The fidelities are computed with respect to a state encoded in circuit $U$ which is assumed to be already initialized.}
    \label{fig:fidelity}
\end{figure}

The decompositions above use wavefunction syntax for convenience and are only feasible for problems that can be encoded into small analytical wave functions. 
Projectors on larger, potentially unknown, wavefunctions can be encoded as well, as long as the unitary circuit that prepares them is known. The projector then looks like
\begin{align}
    \ket{\Psi}\bra{\Psi} = U_\Psi\ket{0}\bra{0}U_\Psi^\dagger,
    \label{eq:general_projector}
\end{align}
where the unitary $U_\Psi$ is a \tequila quantum circuit. In Fig.~\ref{fig:fidelity} we illustrate both techniques applied for the evaluation of fidelities $F=\lvert\bra{\Psi}\ket{\Phi}\rvert^2$.

\subsubsection{Quantum Circuits}\label{sec:circuits}

Abstract quantum circuits can be defined over primitive quantum gates, either from those in the \tequila gate set, or by defining them over a generator which itself is given by a \tequila Hamiltonian.
As an example, consider the  $y$-rotation on qubit $0$,
\begin{align}
    R_y(a) = e^{-\frac{a}{2} \sigma_y(0)},
\end{align}
which can be initialized in the following equivalent ways:
\begin{lstlisting}[language=Python]
U = tq.gates.Ry(angle=1.0, target=0)
U = tq.gates.ExpPauli(angle=1.0,
    paulistring="Y(0)")
g = tq.paulis.Y(0)
U = tq.gates.Trotterized(angles=[1.0],
    generators=[g], steps=1)
\end{lstlisting}
The \texttt{Trotterized} function accepts arbitrary Hermitian generators and follows the same conventions as the one-qubit rotations with the exponent being $-i\frac{a}{2}$. Note that there is no approximation used in this example, since a single step in the formal trotterization is exact here; however this is not the case for all generators. In general \tequila allows the definition of arbitrary gates over Hermitian generators $G=\sum_{k} c_k \boldsymbol{\sigma}_k$, represented by a weighted sum over $k$ Pauli strings $\boldsymbol{\sigma}_k$, as a formal Trotter expansion with $N$ steps,
\begin{equation}
   U(a) = e^{-i\frac{a}{2} G} = \prod_{n=1}^{N}\prod_{k=1}^{K} e^{-i\frac{a}{2} \frac{c_k}{N} \boldsymbol{\sigma}_k}\label{eq:quantum_gate_over_generator}.
\end{equation}

Parametrized gates like rotations can also be initialized by variables or transformations of variables. Take for example the parametrized rotation:
\begin{align}
    R_y(f(a)) = e^{-i\frac{f(a)}{2}\sigma_y},\;\; f(a) = e^{-\frac{a^2}{4}},
\end{align}
which can be initialized as
\begin{lstlisting}[language=Python]
a = tq.Variable("a")
fa = (-0.25*a**2).apply(exp)
U = tq.gates.Ry(angle=fa, target=0)
\end{lstlisting}
Here \texttt{exp} is the exponential function, for example from \textsc{numpy}, and the transformation of abstract variables uses the same data structures as the abstract objectives.
The actual values of the variables are not tied to the circuit structures but are passed down when the objectives are evaluated (see previous sections). In this way, the objectives can be used like abstract functions.

Quantum circuits usually consist of more than one gate, and their construction can be achieved by simply adding individual gates together. 
The arithmetic is here implemented according to the quantum circuit model where addition is interpreted as concatenating two circuits, and the leftmost circuit is the one that acts first.
Take for example, the following unitary operation consisting of four individual unitary operations written in the opposite order using the language of unitary matrices
\begin{align}
    U &= U_dU_cU_bU_a \nonumber \\
    &= \raisebox{-5pt}{\includegraphics[scale=0.7]{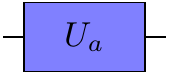}} + \raisebox{-5pt}{\includegraphics[scale=0.7]{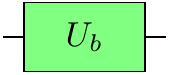}} + \raisebox{-5pt}{\includegraphics[scale=0.7]{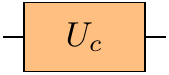}} + \raisebox{-5pt}{\includegraphics[scale=0.7]{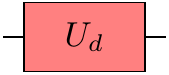}} \nonumber\\
    & =\raisebox{-5pt}{\includegraphics[scale=0.7]{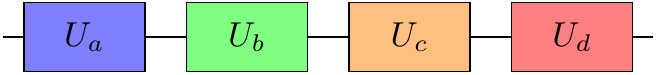}}.
\end{align}
At the top level, \tequila does not distinguish between gates and circuits. The rotations above are initialized as circuits containing a single gate. See, for example, Fig.~\ref{fig:example_all_in_one}, and the applications illustrated later.

Circuits can be evaluated in the same way as objectives if one is interested in simulating the wave function or sampling from its distribution and similarly with the \texttt{compile} function:
\begin{lstlisting}[language=Python]
wfn = tq.simulate(U, variables={"a":1.0})
\end{lstlisting}
The returned data type is the same for finite sampling (\texttt{samples=finite\_value)} and explicit simulation (\texttt{samples=None}), where the former contains the counts for the corresponding measurements and the latter the simulated amplitudes.

\subsection{Simulation and execution}\label{sec:simulation_execution}

Some improvements in algorithmic quantum computing concern more technical details in its implementation such as efficient unitary gate compilation, error mitigation or optimizing measurement protocols. 
In these cases, it is not sufficient to treat the quantum computer as a black box, and more details about the device are necessary.
\tequila, due to its modular design, offers the platform to implement and incorporate these technical improvements quickly and easily, making them available, accessible, and easy to use for a broad user base.
For instance, a feature already included in \tequila is the automatic compilation of abstract multi-qubit gates (like multi-Pauli rotations and controlled-rotations) into primitive quantum gates and the option to optimize measurement protocols by grouping the Hamiltonian into commuting cliques.

\subsubsection{Optimized Measurement Protocols}

The protocol for grouping a Hamiltionian into commuting and qubit-wise commuting cliques are described in detail in Refs. \cite{yen2020} and \cite{verteletskyi2020}, respectively.  
The implementation thereof uses the binary-symplectic representation of the Hamiltonian, where each Pauli string is represented by two integer arrays (see \cite{yen2020} for more details). \tequila can convert the standard Hamiltonians into binary-symplectic form and vice versa. The grouping algorithm transforms each expectation value of an objective into multiple expectation values of the form
\begin{align}
    \expval{H}_U = \sum_i\expval{\tilde{H}_i}_{U_iU}
\end{align}
where the individual Hamiltonians $\tilde{H}_i$ are built up solely from Pauli-$Z$ and unit operators and the $U_i$ correspond to Clifford gates added to the original unitary $U$. 
Optimizing the measurement protocol for an \text{Objective} will return an \texttt{Objective} as well.  After the optimization, all non-trivial operators in the Hamiltonians are Pauli-$Z$ operators, necessitating only a single type of measurement to be performed for each of them. This type is automatically detected and applied if an evaluation with finite sample sizes is requested.
\tequila will then request measurements of the qubits supported in the Hamiltonian in the computational basis from the corresponding quantum backend.
The measurement counts $p_0(i)$, $p_1(i)$, of qubit $i$ in the computational basis states $\ket{0}$ and $\ket{1}$, are then used to estimate the individual expectation values in the objective. Take for example 
\begin{eqnarray}
    E &=& \expval{ a \ Z(1)+ c \ Z(1)Z(2) }_U \nonumber \\
    &=&  a \ \expval{Z(1)}_U + c \ \expval{Z(1)Z(2)}_U \nonumber\\
    &=& a \left(p_0(1) - p_1(1)\right) \nonumber \\
    &&+\  c \ \left( (p_0(1)-p_1(1))(p_0(2)-p_1(2))\right) 
\end{eqnarray}
where $Z = \ket{0}\bra{0} - \ket{1}\bra{1} $. Note that strict equality holds only for an infinite number of samples where $p_x = |\bra{x}U\ket{0}|^2$ is the exact measurement probability.

Comprehension of the details of the implementation of the measurement optimization requires in-depth technical knowledge about measurement on quantum computers, knowledge which the sole user of such a device does usually not require.
This optimization is available to all \tequila users by the inclusion of a simple \texttt{optimize\_measurements=true} keyword statement when initializing an expectation value (see the online tutorials~\cite{tequila} for explicit examples).

\subsubsection{Gate Compilation and Translation}

In order to simulate or execute abstract circuits and objectives, the structures must first be translated into the appropriate backend at the gate level. In \tequila this can be accomplished either at time of execution, or by compilation in advance through the \texttt{tq.compile} function. In its base version, \tequila offers compilation for gradients and backends for (controlled) rotations, (controlled) exponentiated Pauli gates (power gates) and arbitrary quantum gates defined over generators as in Eq.~\eqref{eq:quantum_gate_over_generator}. For the future, we anticipate further improvement by the integration of specialized compiler packages such as \textsc{tket}~\cite{tket} or \textsc{pyzx}~\cite{kissinger2020Pyzx} for which basic functionality is already supported through corresponding interfaces within \tequila.

Evaluation of a quantum circuit requires two protocols: compilation, followed by translation. In compilation, abstract \tequila gates are mapped to a more restricted set of gates by a \tequila compiler object. This object is generally deployed within a pre-defined \tequila function, although compilers may also be constructed by the user. The aforementioned compiler sequentially performs a number of compilations according to a series of Boolean arguments received upon initialization. These compilations can be translation of multi-control or multi-target gates into a sequence of single control or single target gates, compilation of controlled rotations into CNOT and single qubit gates, the compilation of power gates into rotation gates, etc. This compiler is also required when automatic differentiation of objectives is performed, and is handled automatically as a subroutine of \texttt{tq.grad}. 

After an abstract circuit (or \texttt{Objective} containing several abstract circuits) is compiled to a reduced gate set, it may then be translated into the language of a specific quantum backend like \textsc{qulacs}, \textsc{qibo}, \textsc{qiskit}, \textsc{cirq} or \textsc{pyquil}. This is accomplished through the \texttt{BackendCircuit} object, more specifically through backend-specific inheritors of this class, such as \texttt{BackendCircuitCirq} or \texttt{BackendCircuitQulacs}, etc. These classes create the respective circuits in their target packages from \tequila \texttt{QCircuit} objects upon initialization, and handle all tasks of simulation, sampling, updating of variables, etc. for the target quantum backend. Because each backend supports only a subset of the available \tequila abstract gates directly, each \texttt{BackendCircuit} inheritor class contains the hard-coded list of compilation instructions required to map an arbitrary \tequila \texttt{QCircuit} into one containing only operations which are individually translatable into operations supported in the target backend. After the necessary compilation is performed, the \texttt{BackendCircuit} inheritor then translates each operation, differentiating between parametrized gates, unparametrized gates, and measurement operations as it does so in order to map \tequila parameter arguments (variables and objectives thereof) into the variable placeholders used by the quantum backend.

Note that compiled objectives can still be combined in the same fashion as before. This allows having objectives with expectation values that are themselves evaluated on different quantum backends. Therefore applications where different parts of a variational algorithm are executed on different hardware or where part of an algorithm is simulated classically are naturally realizable within \tequila.

\subsubsection{Optimizers}\label{sec:optimizers}

Iterative classical optimization is a core subroutine of any VQA, and \tequila is purposefully tailored toward simplifying this task. Any parametrized tequila objective can be optimized, either through the use of the built-in \tequila gradient descent optimizer, or through a number of optimizer objects that provide an interface between \tequila and powerful optimization packages like  \textsc{SciPy}~\cite{scipy}, \textsc{GPyOpt}~\cite{gpyopt2016}, or \textsc{Phoenics}~\cite{hase2018phoenics}. All optimizers in \tequila inherit from a shared base class, designed for ease of extension. Additionally, \tequila implements a class called \texttt{History}, which allows for easy manipulation and plotting of the trajectories followed by objectives and their parameters over the course of some optimization run. All the optimizers are callables, taking an \texttt{Objective} alongside a variety of keyword arguments, and return a \texttt{NamedTuple} specific to that optimizer. 

All supported methods are conveniently accessible through the \texttt{minimize} function,
\begin{lstlisting}[language=Python]
result = tq.minmize(objective, method="bfgs")
\end{lstlisting}
which can take the same keywords as \texttt{compile} and \texttt{simulate} in order to control the quantum backend and additional optimizer specific keywords like \texttt{gradient}, \texttt{initial\_values}, or \texttt{maxiter}. Partial optimization of a specific set of variables can be achieved by passing them as a list with the \texttt{variables} keyword. 
We refer to Fig.~\ref{fig:example_all_in_one}, the application section below and the tutorials provided on Github~\cite{tequila} for explicit use cases and illustrations.

\tequila has its own gradient descent optimizer, capable of optimizing by a variety of popular optimization routines, such as \textit{Adam}, \textit{RMS-prop}, \textit{Nesterov Momentum}, and more. Those methods can be selected by the \texttt{method} keyword in the \texttt{minimize} function, in the same way as presented in the code snipped above. A detailed overview can be found in the online tutorials.~\cite{tequila} In addition to the usual use through a function call, the \tequila gradient descent optimizer can be used as a step-wise optimizer, to give users a more fine-grained control over its use.
Both the \tequila \texttt{GDOptimizer} and the \texttt{SciPyOptimizer} can accept custom gradients provided by the user over the \texttt{gradient} keyword. These optimizers can combine their classical update routines with the Quantum Natural Gradient (QNG)~\cite{Stokes2020quantumnatural} a method of transforming the gradients of expectation values based on the Fubini-Study metric of said expectation value. Numerical gradients are also available for all optimizers by setting the \texttt{gradient} keyword accordingly within the \texttt{minimize} function:
\begin{lstlisting}[language=Python]
# use SciPy's 2-point scheme
result = tq.minmize(objective, 
            method="bfgs",
            gradient="2-point"
            method_options={"finite_diff_rel_step":1.e-4})
# use tequila's 2-point scheme
result = tq.minmize(objective, 
            method="bfgs",
            gradient={"method":"2-point",
                      "stepsize":1.e-4})
\end{lstlisting}

At present, \tequila has plugins to two Bayesian optimization packages: \textsc{GPyOpt}~\cite{gpyopt2016},  and \textsc{Phoenics}~\cite{hase2018phoenics}. These packages allow for robust global optimization and may serve well for objectives that contain a small or intermediate number of parameters but whose gradients require a large array of expectation values. Bayesian optimization has shown promise in the optimization of quantum circuits.~\cite{zhu2019classifier} These optimizers can be accessed identically to the gradient descent optimizers, chiefly through the \texttt{tq.minimize} function. See, for example, Fig.~\ref{fig:example_all_in_one}.

\subsubsection{Noise} \label{sec:noise}

Because of the noise-prone nature of near term quantum devices, the exploration of how a VQA behaves in the presence of quantum noise is crucial to evaluate the performance of the algorithm at hand. Because the formalism for the simulation of quantum noise varies considerably among quantum backends, \tequila attempts to abstract away more pain-staking details of noisy simulation so as to allow comparison between multiple backends. A few assumptions are made in order to accomplish this task:
\begin{enumerate}
    \item All operations can be affected by noise. 
    \item Individual noise sources are independent of each other. 
    \item Each noise source affects all \textit{k}-qubit gates equally. 
    \item Each noise source affects \textit{k}-qubit gates independently from 1,\dots \textit{k-1}, \textit{k+1}, \dots \textit{n}-qubit gates (i.e, 1-qubit gates are noised seperately from 2-qubit gates, etc). 
    \item The effects of noise on a given operation are independent from its position in the circuit. 
\end{enumerate}
%Among these assumptions, the third is perhaps most questionable, but from the perspective of implementation, 
These assumptions are currently necessary in order to make the quantum noise in \tequila backend-independent. Future releases of \tequila may permit more fine-grained, backend-restricted control over quantum noise; only universally-supported operations have been incorporated in this first release.

In \tequila, noise is represented by a \texttt{NoiseModel} object, itself a container for various \texttt{QuantumNoise} objects. Six types of quantum noise are supported in \tequila , these being: bit flips, phase flips, amplitude damps, phase damps, simultaneous amplitude-phase damps, and symmetric depolarizations. Each \texttt{QuantumNoise} object holds a probability (or list thereof), and a level, designating the number of qubits in the operations it should act upon. The \texttt{NoiseModel} object contains and wraps over these \texttt{QuantumNoises}, and is passed to a quantum backend upon translation of a \texttt{QCircuit} or \texttt{Objective}.  \texttt{NoiseModel} objects may be combined with each other through simple addition. 

The manner in which a \texttt{NoiseModel} object is translated into the application of noise depends entirely on the quantum backend in question.  \texttt{NoiseModel} objects may be passed to \texttt{simulate}, \texttt{compile}, and \texttt{minimize} through the \texttt{noise} keyword argument of each function. Note that, because noise is probabilistic, the application of noise requires sampling; it can currently not be combined with wavefunction simulation.

Shown below is a simple example of the construction of a \texttt{NoiseModel} and its application to the simulation of a simple quantum expectation value
\begin{lstlisting}[language=Python]
U = tq.gates.X(0) + tq.gates.CNOT(0,1)
H = tq.Paulis.Qm(1)
O = tq.ExpectationValue(U,H)
bf_1 = tq.circuit.noise.BitFlip(p=0.1,level=1)
bf_2 = tq.circuit.noise.BitFlip(p=0.3,level=2)
my_noise_model = bf_1 + bf_2
E_noisy=tq.simulate(O,samples=5000,noise=my_noise_model)
\end{lstlisting}
where the noise model instructs the quantum backend to apply \texttt{BitFlip} noise with probability $p=0.1$ to all single-qubit gates and $p=0.3$ to all two-qubit gates.

\subsubsection{Real devices}

\tequila is capable of executing and emulating  circuits and objectives on real quantum devices through their corresponding interfaces: The quantum backends. To access or emulate a device, the user only needs to supply a keyword argument, \texttt{device}, to \texttt{compile}, \texttt{simulate}, or \texttt{minimize}. 
The value of this keyword can be any of several types, depending on the quantum backend through which the device is accessed; in general, this may be a string, a dictionary, or some backend-specific instantiation of the device as an object in itself. 

In the case of device emulation, the user can include the specific properties and gate set of a real device in the simulation by setting the device-specific noise model in \texttt{noise='device'}, provided that the quantum backend in question provides some means of access to this noise modeling, as do, for example, \textsc{Qiskit} and \textsc{PyQuil}. 

Similar to noise modeling, the use of a device requires sampling. Shown below is a small example of how a \tequila user with an \textsc{IBMQ} account may use \tequila to run a circuit on the IBMQ Vigo device
\begin{lstlisting}[language=Python]
from qiskit import IBMQ
IBMQ.load_account()
U = tq.gates.Rx('a',0)
H = tq.Paulis.Y(0)
O = tq.ExpectationValue(U,H)
E = tq.simulate(O,backend='qiskit',
                device='ibmq_vigo',
                samples=1000,
                variables={'a':np.pi/2})
\end{lstlisting}
where more fine-grained device specifications, e.g. including \textsc{IBMQ} providers, can be specified by passing either a dictionary or an already initialized \textsc{IBMQ} device.

\section{Examples and Applications}

\tequila is a general purpose library and aims  to simplify initial prototyping, testing, and deployment for quantum algorithm development.
It currently provides extended features for quantum chemistry applications~\cite{kottmann2020reducing} and has also been used in the context of quantum optics~\cite{kottmann2020quantum} and VQE extensions~\cite{cerveralierta2020metavariational}.
In the following, we will illustrate how \tequila can be applied through explicit application examples.
Further, more detailed illustrations can be found in tutorials provided on Github~\cite{tequila}.

\subsection{Quantum Chemistry}

One of the proposed \textit{killer applications}~\cite{reiher2017elucidating} for quantum computers, and the original application proposed for the quantum variational eigensolver~\cite{peruzzo2014variational, McClean2016theoryofvqe} is the electronic structure problem of quantum chemistry.
The goal of this application is to find well-behaved approximations to the eigenvalues of the electronic Hamiltonian which describes the electronic energy of molecular systems within the Born-Oppenheimer approximation.

The electronic Hamiltonian for a molecule with $N_{e}$ electrons can be written as
\begin{align}
    H\left(\Vec{r}_1, \dots, \Vec{r}_{N_{\text{e}}}\right) = \sum_k^{N_\text{e}}h\left(\Vec{r}_k\right) + \frac{1}{2}\sum_{k\neq l}^{N_\text{e}} g\left(\Vec{r}_k, \Vec{r}_l\right),\label{eq:H_electronic}
\end{align}
where the one-electron potential $h=T+V$ is the combination of the one-electron kinetic energy operator $T$ and the summed Coulomb potentials $V$ between a single electron and the nuclear point charges, and $g\left(\Vec{r}_k, \Vec{r}_l\right)$ is the electron-electron Coulomb potential between two electrons at positions $\Vec{r}_k$ and $\Vec{r}_l$ .
Given a set of orthonormal one-electron basis-functions (spin-orbitals) the electronic Hamiltonian can be written in second-quantized form as
\begin{align}
    H = \sum_{kl}h_{kl} a_k^\dagger a_l + \frac{1}{2}\sum_{klmn} g_{klmn} a_k^\dagger a^\dagger_l a_n a_m,\label{eq:H_electronic_second_quantized}
\end{align}
where $a_k^\dagger$ ($a_k$) are anti-commuting operators that create (annihilate) electrons in the spin-orbital $k$ and $h$,$g$ denote the integrals of the corresponding operators and spin-orbitals in Dirac notation. If the set of spin-orbitals forms a complete orthonormal system, the second quantized Hamiltonian~\eqref{eq:H_electronic_second_quantized} is equivalent to the electronic Hamiltonian~\eqref{eq:H_electronic} restricted to the anti-symmetric wave functions obeyed by fermionic systems.
The fermionic anti-symmetry of the electronic states is conveniently guaranteed through the anti-commutation relations of the second-quantized operators. Detailed introductions can be found in textbooks~\cite{helgaker2014molecular, shavitt2009many, jorgensen2012second, surjan2012second},  and in recent reviews~\cite{cao2019quantumreview, mcardle2020quantumreview, fermann2020fundamentals, nisq_review}. 

The second-quantized Hamiltonian can be encoded into a qubit Hamiltonian by application of the Jordan-Wigner or Bravyi-Kitaev transformations~\cite{bravyi2002, seeley2012}, or more recently developed encodings such as the Bravyi-Kitaev Superfast transformation~\cite{setia2018}.
\tequila provides a convenient interface to initialize qubit encoded electronic Hamiltonians by deploying the transformations implemented in \textsc{openfermion}~\cite{openfermion} and allows convenient integration of user-defined transformations..
The molecular integrals can either be supplied as \textsc{numpy} arrays or may be calculated by interfacing electronic structure packages such as  \textsc{psi4}~\cite{psi42020}.
\tequila initializes a \texttt{Molecule} object which can then initialize qubit encoded electronic Hamiltonians and qubit encoded excitation generators, and can serve as an interface to classical methods of \textsc{psi4}.
The \texttt{Molecule} object ensures that the use of encodings, active-spaces, basis-sets and molecular parameters stays consistent for all further initialization. 

\subsubsection{Molecules and Hamiltonians}

Using only their high-level functionality, a \texttt{Molecule} can be initialized by providing the molecular geometry and the one- and two-electron integrals $h$ and $g$. The constant nuclear-repulsion can optionally be provided to be included in the results.
If \textsc{psi4} is installed, the molecular parameters and integrals can be computed automatically by providing a Gaussian basis set (see for example Ref.~\cite{fermann2020fundamentals}).
Molecular structure data can be initialized explicitly or by passing a string that addresses a file in \textit{xyz} format.

The following short example instructs one upon how to initialize a \texttt{Molecule} with \textsc{psi4} or by providing the integrals as \textsc{numpy} arrays \texttt{h} and \texttt{g}, which are assumed to be already initialized here. 
In order to comply with other quantum computing packages, the electron repulsion integrals $g$ are expected in the \textsc{openfermion} convention. In the example below we illustrate how to construct a \tequila molecule from the \textsc{psi4} interface or manually from \textsc{numpy} arrays of molecular integrals:
\begin{lstlisting}[language=Python]
# Can be filename.xyz or explicit string
geomstring = "He 0.0 0.0 0.0"

# Molecule construction with Psi4
molecule = tq.chemistry.Molecule(
    geometry=geomstring,
    basis_set="6-31G",
    transformation="bravy_kitaev")
\end{lstlisting}
\begin{lstlisting}[language=Python]
# Manual molecule construction
# resort g integrals (given as numpy array) 
# from Mulliken to openfermion convention
g = tq.chemistry.NBodyTensor(elems=g,     scheme='mulliken')
g.reorder(to='openfermion')
g = g.elems
molecule = tq.chemistry.Molecule(
    backend="base",
    geometry=geomstring,
    one_body_integrals=h,
    two_body_integrals=g,
    nuclear_repulsion=0.0,
    transformation="bravy_kitaev")
\end{lstlisting}
The two-body integrals $g$ were assumed to be in Mulliken notation ($g_{prqs}^{\text{Mulliken}} \equiv (pq|g|rs) \equiv \bra{pr}g\ket{qs} \equiv g_{prqs}^{Dirac}$), used for example, by \textsc{psi4}.
If installed, \textsc{psi4} is automatically detected by \textsc{tequila} and used as a default.
The \texttt{backend} keyword allows to demand specific backends if multiple supported backends are installed.
Most functionalities are implemented in a backend independent manner, allowing for convenient introduction of novel ways to represent Hamiltonians, such as for example a basis-set-free approach~\cite{kottmann2020reducing} representing the Hamiltonian with directly determined pair-natural orbitals~\cite{kottmann2020direct}.

The \texttt{transformation} keyword specifies the qubit encoding that will be used for further operations, i.e Jordan-Wigner, etc. Currently, all transformations within \textsc{openfermion} are supported. Alternatively, a callable object can be passed as \texttt{transformation} allowing easy integration of user-defined transformations, providing a convenient interface for the future integration of recently developed fermion-to-qubit transformations.~\cite{setia2018, chien2020custom, derby2020compact} Additional parameters for qubit encodings, as for exampl needed within the qubit-tapered~\cite{bravyi2017tapering} form of the Bravyi-Kitaev transformation can be passed down as keyword arguments using \texttt{transformation\_\_} as prefix. If not provided \textsc{tequila} will try to guess them. We refer to the online tutorials~\cite{tequila} for explicit examples.

\subsubsection{Active Spaces}

If the \textsc{psi4} interface is used, active spaces can be set through the \texttt{active\_orbitals} keyword by selecting orbitals which are labeled by their irreducible representation for the underlying point group of the molecule. Additionally, the occupation of the \texttt{reference} determinant can be defined, where the default is the Hartree-Fock reference computed by \textsc{psi4} (the determinant constructed by the first $\frac{N_\text{e}}{2}$ spatial orbitals). Custom reference orbitals can be chosen with the \texttt{reference} keyword using the same input format as for the \texttt{active\_orbitals}. Without \textsc{psi4} as a backend, irreducible representations are not considered and active spaces are set by an array of indices representing the active orbitals.

Active spaces are tied to the \texttt{Molecule} objects and shouldn't be changed after initialization. When the active space is set, all Hamiltonians, excitation generators, classical amplitudes and energies are computed within that active space.
The following example illustrates the initialization of a benzene molecule, restricted to the active space of its 6 conjugated $\pi$ orbitals (three occupied and three unoccupied):
\begin{lstlisting}[language=Python]
active = {
    "B1u": [0,1], 
    "B3g": [0,1], 
    "B2g": [0],
    "Au": [0],
}
molecule = tq.chemistry.Molecule(
    geometry="benzene.xyz",
    basis_set='sto-3g',
    active_orbitals=active
)
\end{lstlisting}
Note that \textsc{psi4} does not support the full $\text{D}_\text{6h}$ point-group and uses $\text{D}_\text{2h}$ instead, leading to different irreducible representations for the degenerate $\pi$ orbitals ($\text{A}_\text{2u} \rightarrow \text{B}_\text{1u}$, $\text{E}_\text{1g} \rightarrow \text{B}_\text{2g}, \text{B}_\text{3g}$, $\text{E}_\text{2u} \rightarrow \text{B}_\text{1u}, \text{A}_\text{u}$, $\text{B}_\text{2g} \rightarrow \text{B}_\text{3g}$).

\subsubsection{Unitary Coupled-Cluster}

Unitary coupled-cluster (UCC) has become a promising model for quantum chemistry on quantum computers~\cite{peruzzo2014variational, McClean2016theoryofvqe} and several promising extensions thereof have been developed in recent years.
Examples include extended strategies~\cite{romero2018strategies, gard2020efficient, yalouz2020stateaveraged, sokolov2020} pair-excitation based~\cite{lee2018generalized} and adaptive strategies in the qubit~\cite{ryabinkin2018qubit, ryabinkin2020iterative, lang2020iterative} or fermionic~\cite{grimsley2019adaptive, tang2019qubit} representation.
\tequila allows the user to combine unitary operators in the UCC framework to develop new approaches.

The basic building blocks are unitary operators,
\begin{align}
    U_{\mathbf{p}\mathbf{q}}\left(a\right) = e^{-i\frac{a}{2} G_{\mathbf{p}\mathbf{q}}},
\end{align}
generated by the hermitian fermionic $n$-body excitation generators
\begin{align}
    G_{pq} &= i(a_p^\dagger a_q - a^\dagger_q a_p)\label{eq:UCC_generator_singles}\\
    G_{pqrs} &= i(a_p^\dagger a_q a_r^\dagger a_s - h.c.)\label{eq:UCC_generator_doubles} \\
    G_{\mathbf{p}\mathbf{q}} &= i( \prod_n a_{p_n}^\dagger a_{q_n} - h.c.), \label{eq:UCC_generator} 
\end{align}
where $p,q,r,s$ are arbitrary spin-orbital indices. Qubit encoded generators of this form can be created from the \texttt{Molecule} object by passing a list of $(p_0,q_1)\dots$ tuples to the function \texttt{make\_excitation\_generator} which will return a \tequila Hamiltonian representing the qubit encoded hermitian generators of Eq.~\eqref{eq:UCC_generator}.

The generators can be used to define a \texttt{Trotterized} unitary quantum gate,
\begin{lstlisting}[language=Python]
idx = [(p0,q0),(p1,q1)]

# definition over generators
G = molecule.make_excitation_generator(idx)
U = tq.gates.Trotterized(
    generators=[G],
    angles=["theta"]
    steps=1
    )

# same unitary
U = molecule.make_excitation_gate(idx)
\end{lstlisting}
where a single Trotter step suffices in this case due to the commutativity of the Pauli strings originating from a single excitation generator.~\cite{romero2018strategies}
Both ways of initializing an elementary fermionic excitation unitary result in the same quantum circuits, initialization over the \texttt{make\_excitation\_gate} will, however, provide necessary information for efficient gradient compiling. Here \tequila exploits certain properties of the fermionic generators in order to reduce the runtime of their gradients significantly. Explicit details on this technique can be found in Ref.~\cite{kottmann2020feasible}.

\tequila uses alternating enumeration for spin-orbitals, meaning that the spin-up orbital $p_{\uparrow}$ of spatial orbital $p$ is enumerated with $2p$ and the corresponding spin-down orbital $p_{\downarrow}$ is enumerated as $2p+1$. Note that this enumeration is independent of the chosen qubit encoding.

\subsubsection{Interface to Classical Methods}

\tequila offers a convenient interface to \textsc{psi4}'s various classical methods by calling the \texttt{compute\_energy} method of the \texttt{Molecule}.
Within unitary coupled-cluster the amplitudes of canonical coupled-cluster often come in handy as potential starting points for further optimization, or for screening purposes.
These can be computed with the method \texttt{compute\_amplitudes}.
We refer to the online tutorials for explicit examples.

\subsubsection{Example: 1-UpCCGSD}

As an explicit example we illustrate how to implement the 1-UpCCGSD ansatz of Ref.~\cite{lee2018generalized} for a Hydrogen-Flouride (HF) molecule in an active space with \tequila.
Other molecules and active spaces may be explored in the same fashion by simply replacing the corresponding lines of code.

The UpCCGSD ansatz is built up from the single and double excitation generators of equations~\eqref{eq:UCC_generator_singles} and~\eqref{eq:UCC_generator_doubles} where the doubles are restricted to pairs of doubly-occupied orbitals $G_{p_{\uparrow}q_{\uparrow}p_{\downarrow}q_{\downarrow}}$.
This example employs a single Trotter step, and orders the fermionic operators by orbital number, but note that other orderings are also possible.~\cite{grimsley2019trotterized, Izmaylov2020ontheorder}
The ansatz herein is constructed explicitly, but note that the molecule structure already offers a convenient initialization of the $k$-UpCCGSD unitary through the \texttt{make\_upccgsd\_ansatz} function.
The full code to optimize the UpCCGSD expectation value is as follows:
\begin{lstlisting}[language=Python]
import tequila as tq
# define the active orbitals
active = active = {"A1":[1,2,3,4,5,6], 
    "B1":[0], 
    "B2":[0]}
geomstring = "F 0.0 0.0 0.0\nH 0.0 0.0 1.0"
# initialize the tequila molecule
mol = tq.chemistry.Molecule(
    geometry=geomstring,
    basis_set='sto-3g',
    active_orbitals=active)

# get some classical reference values
EFCI = mol.compute_energy("fci")
# initialize the hamiltonian
H = mol.make_hamiltonian()

# initialize the circuit for the initial state
U = mol.prepare_reference()
# add the 1-UpCCGSD primitive excitations
for i in range(mol.n_orbitals):
    for a in range(i+1,mol.n_orbitals):
        # electronic pair excitation
        U += mol.make_excitation_gate(
            [(2*i, 2*a),(2*i+1, 2*a+1)])
        # single electron excitation: spin-up
        U += mol.make_excitation_gate(
            [(2*i, 2*a)])
        # single electron excitation: spin-down
        U += mol.make_excitation_gate(
            [(2*i+1, 2*a+1)])
        
# form the abstract expectation value
E = tq.ExpectationValue(H=H, U=U)

# bfgs with numerical gradients
result = tq.minimize(objective=E,method="bfgs")
    
print("Final VQE Energy:", result.energy)
print("FCI Energy      :", EFCI)
\end{lstlisting}
where the \textsc{scipy} implementation of the BFGS optimizer is chosen, and the gradient objectives are automatically compiled according to Ref.~\cite{kottmann2020feasible}.

\subsubsection{Example: Sequential Excited State Solver}

One way for VQAs to optimize bound excited states of a given Hamiltonian is to solve for the ground state and project them out of the Hamiltonian~\cite{lee2018generalized, higgott2019variational} by repeating this procedure sequentially.
After solving for $n$ states $\ket{\Psi_i}$, generated by the unitaries $U_i$, the expectation value of the ansatz $U$ and projected Hamiltonian is given by
\begin{align}
    \expval{\tilde{P}H\tilde{P}}_U 
    &= \sum_{i=1}^n \bra{0}U^\dagger\left( 1-\ket{\Psi_i}\bra{\Psi_i} \right) H U\ket{0} \nonumber\\
    &=  \expval{H} - \sum_i^n E_i \bra{0}U^\dagger U_i\ket{0}\bra{0}U_i^\dagger U\ket{0} \nonumber\\
    &= \expval{H} - \sum_i^n E_i \expval{\boldsymbol{Q}_+}_{U_i^\dagger U}
\end{align}
where we used the idempotency of the projector $\tilde{P} = 1- \ket{\Psi_i}\bra{\Psi_i}$ and assumed $\left[H,\tilde{P}\right]=0$ which is true if the $\ket{\Psi_i}$ are true eigenstates of $H$ and will therefore only hold approximately within most VQAs.~\cite{lee2018generalized}
The $\boldsymbol{Q}_+ = \prod_k Q_+(k)$ operator denotes the projector onto the all-zero qubit state $\ket{0\dots 0}\bra{0 \dots 0}$ making the expectation value $\expval{\boldsymbol{Q}_+}_{U_i^\dagger U}$ equivalent to the squared overlap between the current ansatz $U$ and the previously found state generated by $U_i$.

\begin{figure}
    \centering
    \includegraphics[width=0.4\textwidth]{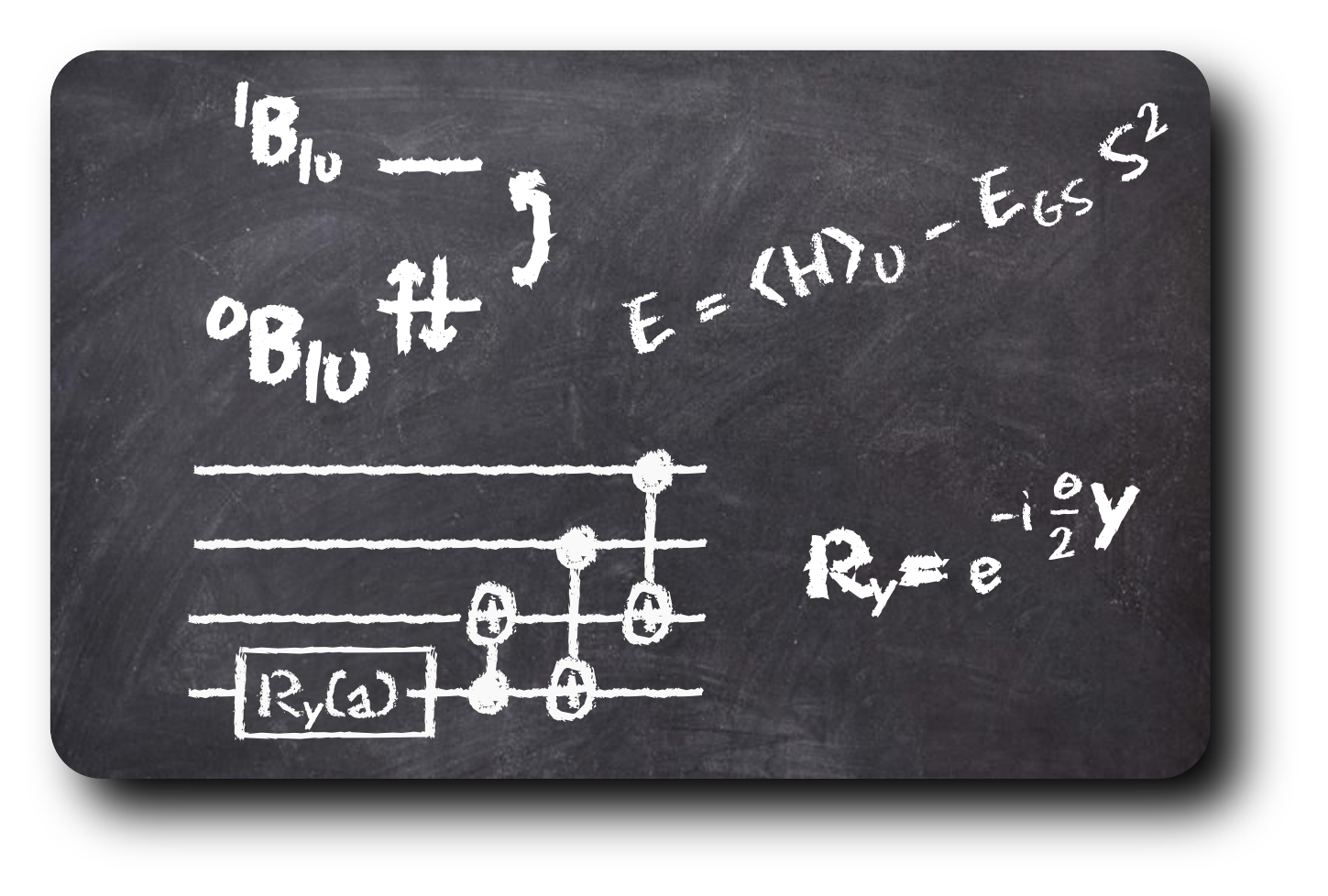}
\begin{lstlisting}[language=Python]
import tequila as tq
# define the geometry directly
geomstring="be 0.0 0.0 0.0\nh 0.0 0.0 {R}\nh 0.0 0.0 -{R}"
R = 1.0
# define the |00><00|=Qp(0)Qp(1) qubit projector
Qp = tq.paulis.Qp([0,1])
# define an active space
active = {"b1u": [0], "b2u": [0]}
# initialize the current molecule
mol = tq.chemistry.Molecule(geometry=geomstring.format(R=R), basis_set="6-31g", active_orbitals=active)
# get the hamiltonian
H = mol.make_hamiltonian()
# collect results 
results = []
# loop over ground and 
# first excited state
for i in range(2):
    # toy circuit
    U = tq.gates.Ry((i, "a"), 0)
    U += tq.gates.CNOT(0, 1) 
    U += tq.gates.CNOT(0, 2)
    U += tq.gates.CNOT(1, 3) 
    U += tq.gates.X([2, 3])
    # initialize an expectation value
    E = tq.ExpectationValue(U, H)
    # get the active variables
    # (keep the old variables static later)
    active_vars = E.extract_variables()
    # initialize the new variables
    angles = {angle: 0.0 for angle in active_vars}
    # Define and add the
    # overlap expectation values
    # from the previous optimizations
    for data, U2 in results:
        S2 = tq.ExpectationValue(H=Qp, 
             U=U2.dagger() + U)
        E -= data.energy * S2
        angles = {**angles, **data.angles}
    # optimize the objective
    result = tq.minimize(E,
        method="bfgs",
        variables=active_vars,
        initial_values=angles)
    # keep the result of the
    # current optimization
    results.append((result, U))
\end{lstlisting}        
    \caption{Toy example of an excited state calculation using the BeH$_2$ molecule in an active space. The top panel illustrates the toy model and the circuit deployed. The bottom panel shows the full implementation with \tequila.}
    \label{fig:example_chemistry_excited_state}
\end{figure}

In Fig.~\ref{fig:example_chemistry_excited_state} we illustrate how such a sequential strategy for excited states can be applied with \tequila. 
In this example we use a customized designed quantum circuit $U$ which could be replaced or combined with
\begin{lstlisting}[language=Python]
U = mol.make_upgccsd_ansatz(label=i)
\end{lstlisting}
to solve with $1$-UpCCGSD as done in Ref.~\cite{lee2018generalized}. Note that the \texttt{label} is added to the variables of the circuit in order to keep the different runs in the sequential solver distinguishable.

\subsection{Variational Quantum Classifier}

\begin{figure}[t!]
    \centering
    \includegraphics[width=0.4\textwidth]{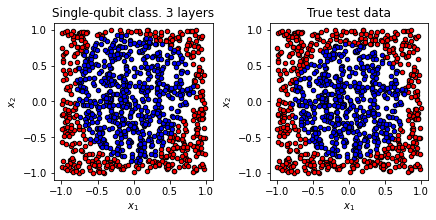}
    \caption{Results after running a 3-layer single-qubit classifier using \tequila. The optimizer used is {RMS-prop} with 400 and 1000 training and test points respectively. The accuracy achieved is 90.5\%.}
    \label{fig:QClass}
\end{figure}

As an example of a quantum machine learning application, consider a Variational Quantum Classifier (VQC). Herein is presented a \tequila tutorial demonstrating the implementation a single-qubit classifier with data re-uploading \cite{QClass}. 
This VQC model encodes data points into single-qubit rotational gates multiple times within the circuit. Each layer is defined as
\begin{equation}
    L\left(\vec{x};\vec{\theta}_{i}\right) = R_{z}\left(x^{1}+\theta_{i}^{1}\right) R_{y}\left(x^{0}+\theta_{i}^{0}\right),
\end{equation}
where $\vec{x}=(x^{0},x^{1})$ are data points and $\vec{\theta}_{i}$ are the optimization parameters of layer $i$. The circuit is a concatenation of layers, similar to other VQAs:
\begin{equation}
    U_{class}\left(\vec{x};\vec{\theta}_{1},...,\vec{\theta}_{l}\right) = L\left(\vec{x};\vec{\theta}_{1}\right)\cdots L\left(\vec{x};\vec{\theta}_{l}\right).
\end{equation}
The single-qubit classifier defines a target state on the Bloch sphere for each class. For a binary classification, the target states are simply the $\ket{0}$ and $\ket{1}$ states.

This circuit definition can be constructed in \tequila using the following function, which depends on the number of layers and the data points:
\begin{lstlisting}[language=Python]
def UClass(x, l)
    U = tq.QCircuit()
        for i in range(l):
            th0 = tq.Variable((0,i))
            th1 = tq.Variable((1,i))
            U += tq.gates.Rx(x[0]+th0, 0)
            U += tq.gates.Ry(x[1]+th1, 1)
    return 
\end{lstlisting}
where abstract variables $\theta$ and numerical values $x$ can be conveniently combined in the initialization of the parametrized gates.

We define the cost function as an objective to be minimized using \tequila optimizers. This \texttt{Objective} is constructed from a training set containing points $\Vec{x}_i$ and corresponding target states $\ket{y_i} \in \left\{ \ket{0}, \ket{1} \right\}$ as
\begin{align}
    L = \sum \left( 1 - F(\Vec{x}_i,y_i) \right)^2,
\end{align}
with the fidelities between training points defined as
\begin{align}
    F(\Vec{x}_i,y_i) &= |\bra{y_i}U_{class}\left(\Vec{x}_i;\Vec{\theta}\right)\ket{0}|^2 \nonumber\\
    &= \expval{P_{y_i}}_{U_{class}(\Vec{x}_i;\Vec{\theta})},
\end{align}
and $P_{y_i}$ as $\ket{y_i}\bra{y_i}$. This is similar to the illustration in Fig.~\ref{fig:fidelity}.
The cost function for this VQC can be initialized as
\begin{lstlisting}[language=Python]
def cost(x, y, l):
    loss = 0.0
    for i in range(len(y)):
        U = UClass(x[i], l)
        ystate ="|{}>".format(y[i])
        P = tq.paulis.Projector(ystate)
        F = tq.ExpectationValue(H=P, U=U)
        loss = loss + (1 - F)**2
    return loss / len(x)
\end{lstlisting}
and, after initialization, can be optimized with one of the optimizers provided by \tequila.

The results of a 2-D circle classification problem are shown in Fig. \ref{fig:QClass}. Using a 3-layer single-qubit classifier, 400 training points and the {RMS-prop} optimization algorithm, the accuracy achieved is in this case 90.5\%.\\

\section{Conclusion}

Herein we have introduced \tequila, a full-stack open-source \textsc{python} package for the rapid development and deployment of variational quantum algorithms.  Through the deployment of novel callable structures, the incorporation of automatic differentiation, and the inclusion of extensible plugins for numerical optimization, quantum chemistry, and more, \tequila is primed for the convenient and intuitive transformation of ideas into code. We seek to continuously forge \tequila into a wide and robust platform, permitting the quantum community to work in a shared and accessible framework to further embolden the ever-accelerating pace of quantum information science and quantum learning, in the hope of enhancing collective mastery of the tools set to emerge in the coming era.\\

\section*{Acknowledgements}

We thank Shumpei Kobayashi for improvements on the optimizer callbacks and Olga Orkut, Hermanni Heimonen, Paul Nation, and Micha\l{} St\c{e}ch\l{}y for valuable feedback.
This work was supported by the U.S. Department of Energy under Award No. DE-SC0019374 as well as DE-AC02-05CH11231 through LBNL subgrant 505736.
A.A.-G. acknowledges the generous support from Google, Inc.  in the form of a Google Focused Award. A.A.-G. also acknowledges support from the Canada Industrial Research Chairs  Program and the Canada 150 Research Chairs Program. 
A.A-G. also acknowledges the Vannevar Bush Faculty Fellowship under contract ONR N00014-16-1-200.
P.S. acknowledges support by a fellowship within the IFI programme of the German Academic Exchange Service (DAAD).
A.F.I. acknowledges financial support from Zapata Computing Inc. and 
the Google Quantum Research Program.
Computations were performed on the niagara supercomputer at the SciNet HPC Consortium.~\cite{niagara1, niagara2} SciNet is funded by: the Canada Foundation for Innovation; the Government of Ontario; Ontario Research Fund - Research Excellence; and the University of Toronto.
S.C. acknowledges support from the MITACS globalink program.
Contributions from B.S., G.T, and C.Z.-M. were made within the QC Mentorship Program of the Quantum Open Source Foundation.  
We thank the generous support of Anders G. Fr\o{}seth.

\newpage
\bibliography{main}

\begin{thebibliography}{74}%
\makeatletter
\providecommand \@ifxundefined [1]{%
 \@ifx{#1\undefined}
}%
\providecommand \@ifnum [1]{%
 \ifnum #1\expandafter \@firstoftwo
 \else \expandafter \@secondoftwo
 \fi
}%
\providecommand \@ifx [1]{%
 \ifx #1\expandafter \@firstoftwo
 \else \expandafter \@secondoftwo
 \fi
}%
\providecommand \natexlab [1]{#1}%
\providecommand \enquote  [1]{``#1''}%
\providecommand \bibnamefont  [1]{#1}%
\providecommand \bibfnamefont [1]{#1}%
\providecommand \citenamefont [1]{#1}%
\providecommand \href@noop [0]{\@secondoftwo}%
\providecommand \href [0]{\begingroup \@sanitize@url \@href}%
\providecommand \@href[1]{\@@startlink{#1}\@@href}%
\providecommand \@@href[1]{\endgroup#1\@@endlink}%
\providecommand \@sanitize@url [0]{\catcode `\\12\catcode `\$12\catcode
  `\&12\catcode `\#12\catcode `\^12\catcode `\_12\catcode `\%12\relax}%
\providecommand \@@startlink[1]{}%
\providecommand \@@endlink[0]{}%
\providecommand \url  [0]{\begingroup\@sanitize@url \@url }%
\providecommand \@url [1]{\endgroup\@href {#1}{\urlprefix }}%
\providecommand \urlprefix  [0]{URL }%
\providecommand \Eprint [0]{\href }%
\providecommand \doibase [0]{https://doi.org/}%
\providecommand \selectlanguage [0]{\@gobble}%
\providecommand \bibinfo  [0]{\@secondoftwo}%
\providecommand \bibfield  [0]{\@secondoftwo}%
\providecommand \translation [1]{[#1]}%
\providecommand \BibitemOpen [0]{}%
\providecommand \bibitemStop [0]{}%
\providecommand \bibitemNoStop [0]{.\EOS\space}%
\providecommand \EOS [0]{\spacefactor3000\relax}%
\providecommand \BibitemShut  [1]{\csname bibitem#1\endcsname}%
\let\auto@bib@innerbib\@empty
%</preamble>
\bibitem [{\citenamefont {Preskill}(2018)}]{preskill2018nisq}%
  \BibitemOpen
  \bibfield  {author} {\bibinfo {author} {\bibfnamefont {J.}~\bibnamefont
  {Preskill}},\ }\bibfield  {title} {\bibinfo {title} {Quantum computing in the
  {NISQ} era and beyond},\ }\href {https://doi.org/10.22331/q-2018-08-06-79}
  {\bibfield  {journal} {\bibinfo  {journal} {Quantum}\ }\textbf {\bibinfo
  {volume} {2}},\ \bibinfo {pages} {79} (\bibinfo {year} {2018})}\BibitemShut
  {NoStop}%
\bibitem [{\citenamefont {Bharti}\ \emph {et~al.}(2021)\citenamefont {Bharti},
  \citenamefont {Cervera-Lierta}, \citenamefont {Kyaw}, \citenamefont {Haug},
  \citenamefont {Alperin-Lea}, \citenamefont {Anand}, \citenamefont {Degroote},
  \citenamefont {Heimonen}, \citenamefont {Kottmann}, \citenamefont {Menke}
  \emph {et~al.}}]{nisq_review}%
  \BibitemOpen
  \bibfield  {author} {\bibinfo {author} {\bibfnamefont {K.}~\bibnamefont
  {Bharti}}, \bibinfo {author} {\bibfnamefont {A.}~\bibnamefont
  {Cervera-Lierta}}, \bibinfo {author} {\bibfnamefont {T.~H.}\ \bibnamefont
  {Kyaw}}, \bibinfo {author} {\bibfnamefont {T.}~\bibnamefont {Haug}}, \bibinfo
  {author} {\bibfnamefont {S.}~\bibnamefont {Alperin-Lea}}, \bibinfo {author}
  {\bibfnamefont {A.}~\bibnamefont {Anand}}, \bibinfo {author} {\bibfnamefont
  {M.}~\bibnamefont {Degroote}}, \bibinfo {author} {\bibfnamefont
  {H.}~\bibnamefont {Heimonen}}, \bibinfo {author} {\bibfnamefont {J.~S.}\
  \bibnamefont {Kottmann}}, \bibinfo {author} {\bibfnamefont {T.}~\bibnamefont
  {Menke}}, \emph {et~al.},\ }\bibfield  {title} {\bibinfo {title} {Noisy
  intermediate-scale quantum (nisq) algorithms},\ }\href
  {https://arxiv.org/abs/2101.08448} {\bibfield  {journal} {\bibinfo  {journal}
  {arXiv:2101.08448 [quant-ph]}\ } (\bibinfo {year} {2021})}\BibitemShut
  {NoStop}%
\bibitem [{\citenamefont {Peruzzo}\ \emph {et~al.}(2014)\citenamefont
  {Peruzzo}, \citenamefont {McClean}, \citenamefont {Shadbolt}, \citenamefont
  {Yung}, \citenamefont {Zhou}, \citenamefont {Love}, \citenamefont
  {Aspuru-Guzik},\ and\ \citenamefont {O’brien}}]{peruzzo2014variational}%
  \BibitemOpen
  \bibfield  {author} {\bibinfo {author} {\bibfnamefont {A.}~\bibnamefont
  {Peruzzo}}, \bibinfo {author} {\bibfnamefont {J.}~\bibnamefont {McClean}},
  \bibinfo {author} {\bibfnamefont {P.}~\bibnamefont {Shadbolt}}, \bibinfo
  {author} {\bibfnamefont {M.-H.}\ \bibnamefont {Yung}}, \bibinfo {author}
  {\bibfnamefont {X.-Q.}\ \bibnamefont {Zhou}}, \bibinfo {author}
  {\bibfnamefont {P.~J.}\ \bibnamefont {Love}}, \bibinfo {author}
  {\bibfnamefont {A.}~\bibnamefont {Aspuru-Guzik}},\ and\ \bibinfo {author}
  {\bibfnamefont {J.~L.}\ \bibnamefont {O’brien}},\ }\bibfield  {title}
  {\bibinfo {title} {A variational eigenvalue solver on a photonic quantum
  processor},\ }\href {https://doi.org/10.1038/ncomms5213} {\bibfield
  {journal} {\bibinfo  {journal} {Nat. Commun.}\ }\textbf {\bibinfo {volume}
  {5}},\ \bibinfo {pages} {4213} (\bibinfo {year} {2014})}\BibitemShut
  {NoStop}%
\bibitem [{\citenamefont {Farhi}\ \emph {et~al.}(2014)\citenamefont {Farhi},
  \citenamefont {Goldstone},\ and\ \citenamefont {Gutmann}}]{farhi2014qaoa}%
  \BibitemOpen
  \bibfield  {author} {\bibinfo {author} {\bibfnamefont {E.}~\bibnamefont
  {Farhi}}, \bibinfo {author} {\bibfnamefont {J.}~\bibnamefont {Goldstone}},\
  and\ \bibinfo {author} {\bibfnamefont {S.}~\bibnamefont {Gutmann}},\
  }\bibfield  {title} {\bibinfo {title} {A quantum approximate optimization
  algorithm},\ }\href {https://arxiv.org/abs/1411.4028} {\bibfield  {journal}
  {\bibinfo  {journal} {arXiv:1411.4028 [quant-ph]}\ } (\bibinfo {year}
  {2014})}\BibitemShut {NoStop}%
\bibitem [{\citenamefont {Wittek}(2014)}]{wittek2014qml}%
  \BibitemOpen
  \bibfield  {author} {\bibinfo {author} {\bibfnamefont {P.}~\bibnamefont
  {Wittek}},\ }\href
  {https://www.sciencedirect.com/book/9780128009536/quantum-machine-learning}
  {\emph {\bibinfo {title} {Quantum Machine Learning: What Quantum Computing
  Means to Data Mining}}}\ (\bibinfo  {publisher} {Elsevier},\ \bibinfo {year}
  {2014})\BibitemShut {NoStop}%
\bibitem [{\citenamefont {Quantiki}(2020)}]{quantiki}%
  \BibitemOpen
  \bibfield  {author} {\bibinfo {author} {\bibnamefont {Quantiki}},\ }\href
  {https://www.quantiki.org/wiki/list-qc-simulators} {\bibinfo {title} {List of
  qc simulators}} (\bibinfo {year} {2020})\BibitemShut {NoStop}%
\bibitem [{\citenamefont {cirq developers}(2018)}]{cirq}%
  \BibitemOpen
  \bibfield  {author} {\bibinfo {author} {\bibfnamefont {T.}~\bibnamefont {cirq
  developers}},\ }\href {https://cirq.readthedocs.io/en/stable/index.html#}
  {\bibinfo {title} {Cirq: A python framework for creating, editing, and
  invoking noisy intermediate scale quantum circuits}} (\bibinfo {year}
  {2018})\BibitemShut {NoStop}%
\bibitem [{\citenamefont {Abraham}\ \emph {et~al.}(2019)\citenamefont {Abraham}
  \emph {et~al.}}]{Qiskit}%
  \BibitemOpen
  \bibfield  {author} {\bibinfo {author} {\bibfnamefont {H.}~\bibnamefont
  {Abraham}} \emph {et~al.},\ }\href {https://doi.org/10.5281/zenodo.2562110}
  {\bibinfo {title} {Qiskit: An open-source framework for quantum computing}}
  (\bibinfo {year} {2019})\BibitemShut {NoStop}%
\bibitem [{\citenamefont {Microsoft}(2020)}]{q_sharp}%
  \BibitemOpen
  \bibfield  {author} {\bibinfo {author} {\bibnamefont {Microsoft}},\ }\href
  {https://docs.microsoft.com/en-us/quantum/user-guide/?view=qsharp-preview}
  {\bibinfo {title} {The q\# user guide - microsoft quantum}} (\bibinfo {year}
  {2020})\BibitemShut {NoStop}%
\bibitem [{\citenamefont {Smith}\ \emph {et~al.}(2016)\citenamefont {Smith},
  \citenamefont {Curtis},\ and\ \citenamefont {Zeng}}]{smith2016practical}%
  \BibitemOpen
  \bibfield  {author} {\bibinfo {author} {\bibfnamefont {R.~S.}\ \bibnamefont
  {Smith}}, \bibinfo {author} {\bibfnamefont {M.~J.}\ \bibnamefont {Curtis}},\
  and\ \bibinfo {author} {\bibfnamefont {W.~J.}\ \bibnamefont {Zeng}},\
  }\bibfield  {title} {\bibinfo {title} {A practical quantum instruction set
  architecture},\ }\href {https://arxiv.org/abs/1608.03355} {\bibfield
  {journal} {\bibinfo  {journal} {arXiv:1608.03355 [quant-ph]}\ } (\bibinfo
  {year} {2016})}\BibitemShut {NoStop}%
\bibitem [{\citenamefont {Killoran}\ \emph {et~al.}(2019)\citenamefont
  {Killoran}, \citenamefont {Izaac}, \citenamefont {Quesada}, \citenamefont
  {Bergholm}, \citenamefont {Amy},\ and\ \citenamefont
  {Weedbrook}}]{Killoran2019strawberryfields}%
  \BibitemOpen
  \bibfield  {author} {\bibinfo {author} {\bibfnamefont {N.}~\bibnamefont
  {Killoran}}, \bibinfo {author} {\bibfnamefont {J.}~\bibnamefont {Izaac}},
  \bibinfo {author} {\bibfnamefont {N.}~\bibnamefont {Quesada}}, \bibinfo
  {author} {\bibfnamefont {V.}~\bibnamefont {Bergholm}}, \bibinfo {author}
  {\bibfnamefont {M.}~\bibnamefont {Amy}},\ and\ \bibinfo {author}
  {\bibfnamefont {C.}~\bibnamefont {Weedbrook}},\ }\bibfield  {title} {\bibinfo
  {title} {Strawberry {F}ields: {A} {S}oftware {P}latform for {P}hotonic
  {Q}uantum {C}omputing},\ }\href {https://doi.org/10.22331/q-2019-03-11-129}
  {\bibfield  {journal} {\bibinfo  {journal} {{Quantum}}\ }\textbf {\bibinfo
  {volume} {3}},\ \bibinfo {pages} {129} (\bibinfo {year} {2019})}\BibitemShut
  {NoStop}%
\bibitem [{\citenamefont {Bergholm}\ \emph {et~al.}(2018)\citenamefont
  {Bergholm}, \citenamefont {Izaac}, \citenamefont {Schuld}, \citenamefont
  {Gogolin}, \citenamefont {Blank}, \citenamefont {McKiernan},\ and\
  \citenamefont {Killoran}}]{bergholm2018pennylane}%
  \BibitemOpen
  \bibfield  {author} {\bibinfo {author} {\bibfnamefont {V.}~\bibnamefont
  {Bergholm}}, \bibinfo {author} {\bibfnamefont {J.}~\bibnamefont {Izaac}},
  \bibinfo {author} {\bibfnamefont {M.}~\bibnamefont {Schuld}}, \bibinfo
  {author} {\bibfnamefont {C.}~\bibnamefont {Gogolin}}, \bibinfo {author}
  {\bibfnamefont {C.}~\bibnamefont {Blank}}, \bibinfo {author} {\bibfnamefont
  {K.}~\bibnamefont {McKiernan}},\ and\ \bibinfo {author} {\bibfnamefont
  {N.}~\bibnamefont {Killoran}},\ }\bibfield  {title} {\bibinfo {title}
  {Pennylane: Automatic differentiation of hybrid quantum-classical
  computations},\ }\href {https://arxiv.org/abs/1811.04968} {\bibfield
  {journal} {\bibinfo  {journal} {arXiv:1811.04968 [quant-ph]}\ } (\bibinfo
  {year} {2018})}\BibitemShut {NoStop}%
\bibitem [{\citenamefont {McCaskey}\ \emph {et~al.}(2019)\citenamefont
  {McCaskey}, \citenamefont {Lyakh}, \citenamefont {Dumitrescu}, \citenamefont
  {Powers},\ and\ \citenamefont {Humble}}]{alex2019xacc}%
  \BibitemOpen
  \bibfield  {author} {\bibinfo {author} {\bibfnamefont {A.~J.}\ \bibnamefont
  {McCaskey}}, \bibinfo {author} {\bibfnamefont {D.~I.}\ \bibnamefont {Lyakh}},
  \bibinfo {author} {\bibfnamefont {E.~F.}\ \bibnamefont {Dumitrescu}},
  \bibinfo {author} {\bibfnamefont {S.~S.}\ \bibnamefont {Powers}},\ and\
  \bibinfo {author} {\bibfnamefont {T.~S.}\ \bibnamefont {Humble}},\ }\bibfield
   {title} {\bibinfo {title} {{XACC}: A system-level software infrastructure
  for heterogeneous quantum-classical computing},\ }\href
  {https://arxiv.org/abs/1911.02452} {\bibfield  {journal} {\bibinfo  {journal}
  {arXiv:1911.02452 [quant-ph]}\ } (\bibinfo {year} {2019})}\BibitemShut
  {NoStop}%
\bibitem [{\citenamefont {Steiger}\ \emph {et~al.}(2018)\citenamefont
  {Steiger}, \citenamefont {H{\"{a}}ner},\ and\ \citenamefont
  {Troyer}}]{Steiger2018projectqopensource}%
  \BibitemOpen
  \bibfield  {author} {\bibinfo {author} {\bibfnamefont {D.~S.}\ \bibnamefont
  {Steiger}}, \bibinfo {author} {\bibfnamefont {T.}~\bibnamefont
  {H{\"{a}}ner}},\ and\ \bibinfo {author} {\bibfnamefont {M.}~\bibnamefont
  {Troyer}},\ }\bibfield  {title} {\bibinfo {title} {Project{Q}: an open source
  software framework for quantum computing},\ }\href
  {https://doi.org/10.22331/q-2018-01-31-49} {\bibfield  {journal} {\bibinfo
  {journal} {{Quantum}}\ }\textbf {\bibinfo {volume} {2}},\ \bibinfo {pages}
  {49} (\bibinfo {year} {2018})}\BibitemShut {NoStop}%
\bibitem [{\citenamefont {Efthymiou}\ \emph {et~al.}(2020)\citenamefont
  {Efthymiou}, \citenamefont {Ramos-Calderer}, \citenamefont {Bravo-Prieto},
  \citenamefont {Pérez-Salinas}, \citenamefont {García-Martín},
  \citenamefont {Garcia-Saez}, \citenamefont {Latorre},\ and\ \citenamefont
  {Carrazza}}]{efthymiou2020qibo}%
  \BibitemOpen
  \bibfield  {author} {\bibinfo {author} {\bibfnamefont {S.}~\bibnamefont
  {Efthymiou}}, \bibinfo {author} {\bibfnamefont {S.}~\bibnamefont
  {Ramos-Calderer}}, \bibinfo {author} {\bibfnamefont {C.}~\bibnamefont
  {Bravo-Prieto}}, \bibinfo {author} {\bibfnamefont {A.}~\bibnamefont
  {Pérez-Salinas}}, \bibinfo {author} {\bibfnamefont {D.}~\bibnamefont
  {García-Martín}}, \bibinfo {author} {\bibfnamefont {A.}~\bibnamefont
  {Garcia-Saez}}, \bibinfo {author} {\bibfnamefont {J.~I.}\ \bibnamefont
  {Latorre}},\ and\ \bibinfo {author} {\bibfnamefont {S.}~\bibnamefont
  {Carrazza}},\ }\bibfield  {title} {\bibinfo {title} {Qibo: a framework for
  quantum simulation with hardware acceleration},\ }\href
  {https://arxiv.org/abs/2009.01845} {\bibfield  {journal} {\bibinfo  {journal}
  {arXiv:2009.01845[quant-ph]}\ } (\bibinfo {year} {2020})}\BibitemShut
  {NoStop}%
\bibitem [{\citenamefont {Luo}\ \emph {et~al.}(2020)\citenamefont {Luo},
  \citenamefont {Liu}, \citenamefont {Zhang},\ and\ \citenamefont
  {Wang}}]{YaoFramework2019}%
  \BibitemOpen
  \bibfield  {author} {\bibinfo {author} {\bibfnamefont {X.-Z.}\ \bibnamefont
  {Luo}}, \bibinfo {author} {\bibfnamefont {J.-G.}\ \bibnamefont {Liu}},
  \bibinfo {author} {\bibfnamefont {P.}~\bibnamefont {Zhang}},\ and\ \bibinfo
  {author} {\bibfnamefont {L.}~\bibnamefont {Wang}},\ }\bibfield  {title}
  {\bibinfo {title} {{Yao. jl: Extensible, efficient framework for quantum
  algorithm design}},\ }\href {https://doi.org/10.22331/q-2020-10-11-341}
  {\bibfield  {journal} {\bibinfo  {journal} {Quantum}\ }\textbf {\bibinfo
  {volume} {4}},\ \bibinfo {pages} {341} (\bibinfo {year} {2020})}\BibitemShut
  {NoStop}%
\bibitem [{\citenamefont {Nguyen}\ \emph {et~al.}(2021)\citenamefont {Nguyen},
  \citenamefont {Bassman}, \citenamefont {Lyakh}, \citenamefont {McCaskey},
  \citenamefont {Leyton-Ortega}, \citenamefont {Pooser}, \citenamefont
  {Elwasif}, \citenamefont {Humble},\ and\ \citenamefont
  {de~Jong}}]{nguyen2021composable}%
  \BibitemOpen
  \bibfield  {author} {\bibinfo {author} {\bibfnamefont {T.}~\bibnamefont
  {Nguyen}}, \bibinfo {author} {\bibfnamefont {L.}~\bibnamefont {Bassman}},
  \bibinfo {author} {\bibfnamefont {D.}~\bibnamefont {Lyakh}}, \bibinfo
  {author} {\bibfnamefont {A.}~\bibnamefont {McCaskey}}, \bibinfo {author}
  {\bibfnamefont {V.}~\bibnamefont {Leyton-Ortega}}, \bibinfo {author}
  {\bibfnamefont {R.}~\bibnamefont {Pooser}}, \bibinfo {author} {\bibfnamefont
  {W.}~\bibnamefont {Elwasif}}, \bibinfo {author} {\bibfnamefont {T.~S.}\
  \bibnamefont {Humble}},\ and\ \bibinfo {author} {\bibfnamefont {W.~A.}\
  \bibnamefont {de~Jong}},\ }\href@noop {} {\bibinfo {title} {Composable
  programming of hybrid workflows for quantum simulation}} (\bibinfo {year}
  {2021}),\ \Eprint {https://arxiv.org/abs/2101.08151} {arXiv:2101.08151
  [quant-ph]} \BibitemShut {NoStop}%
\bibitem [{\citenamefont {Suzuki}\ \emph {et~al.}(2020)\citenamefont {Suzuki},
  \citenamefont {Kawase}, \citenamefont {Masumura}, \citenamefont {Hiraga},
  \citenamefont {Nakadai}, \citenamefont {Chen}, \citenamefont {Nakanishi},
  \citenamefont {Mitarai}, \citenamefont {Imai}, \citenamefont {Tamiya},
  \citenamefont {Yamamoto}, \citenamefont {Yan}, \citenamefont {Kawakubo},
  \citenamefont {Nakagawa}, \citenamefont {Ibe}, \citenamefont {Zhang},
  \citenamefont {Yamashita}, \citenamefont {Yoshimura}, \citenamefont
  {Hayashi},\ and\ \citenamefont {Fujii}}]{qulacs}%
  \BibitemOpen
  \bibfield  {author} {\bibinfo {author} {\bibfnamefont {Y.}~\bibnamefont
  {Suzuki}}, \bibinfo {author} {\bibfnamefont {Y.}~\bibnamefont {Kawase}},
  \bibinfo {author} {\bibfnamefont {Y.}~\bibnamefont {Masumura}}, \bibinfo
  {author} {\bibfnamefont {Y.}~\bibnamefont {Hiraga}}, \bibinfo {author}
  {\bibfnamefont {M.}~\bibnamefont {Nakadai}}, \bibinfo {author} {\bibfnamefont
  {J.}~\bibnamefont {Chen}}, \bibinfo {author} {\bibfnamefont {K.~M.}\
  \bibnamefont {Nakanishi}}, \bibinfo {author} {\bibfnamefont {K.}~\bibnamefont
  {Mitarai}}, \bibinfo {author} {\bibfnamefont {R.}~\bibnamefont {Imai}},
  \bibinfo {author} {\bibfnamefont {S.}~\bibnamefont {Tamiya}}, \bibinfo
  {author} {\bibfnamefont {T.}~\bibnamefont {Yamamoto}}, \bibinfo {author}
  {\bibfnamefont {T.}~\bibnamefont {Yan}}, \bibinfo {author} {\bibfnamefont
  {T.}~\bibnamefont {Kawakubo}}, \bibinfo {author} {\bibfnamefont {Y.~O.}\
  \bibnamefont {Nakagawa}}, \bibinfo {author} {\bibfnamefont {Y.}~\bibnamefont
  {Ibe}}, \bibinfo {author} {\bibfnamefont {Y.}~\bibnamefont {Zhang}}, \bibinfo
  {author} {\bibfnamefont {H.}~\bibnamefont {Yamashita}}, \bibinfo {author}
  {\bibfnamefont {H.}~\bibnamefont {Yoshimura}}, \bibinfo {author}
  {\bibfnamefont {A.}~\bibnamefont {Hayashi}},\ and\ \bibinfo {author}
  {\bibfnamefont {K.}~\bibnamefont {Fujii}},\ }\bibfield  {title} {\bibinfo
  {title} {Qulacs: a fast and versatile quantum circuit simulator for research
  purpose},\ }\href {https://arxiv.org/abs/2011.13524} {\bibfield  {journal}
  {\bibinfo  {journal} {arXiv:2011.13524 [quant-ph]}\ } (\bibinfo {year}
  {2020})}\BibitemShut {NoStop}%
\bibitem [{\citenamefont {Parrish}\ \emph {et~al.}(2017)\citenamefont
  {Parrish}, \citenamefont {Burns}, \citenamefont {Smith}, \citenamefont
  {Simmonett}, \citenamefont {DePrince~III}, \citenamefont {Hohenstein},
  \citenamefont {Bozkaya}, \citenamefont {Sokolov}, \citenamefont {Di~Remigio},
  \citenamefont {Richard} \emph {et~al.}}]{psi4}%
  \BibitemOpen
  \bibfield  {author} {\bibinfo {author} {\bibfnamefont {R.~M.}\ \bibnamefont
  {Parrish}}, \bibinfo {author} {\bibfnamefont {L.~A.}\ \bibnamefont {Burns}},
  \bibinfo {author} {\bibfnamefont {D.~G.}\ \bibnamefont {Smith}}, \bibinfo
  {author} {\bibfnamefont {A.~C.}\ \bibnamefont {Simmonett}}, \bibinfo {author}
  {\bibfnamefont {A.~E.}\ \bibnamefont {DePrince~III}}, \bibinfo {author}
  {\bibfnamefont {E.~G.}\ \bibnamefont {Hohenstein}}, \bibinfo {author}
  {\bibfnamefont {U.}~\bibnamefont {Bozkaya}}, \bibinfo {author} {\bibfnamefont
  {A.~Y.}\ \bibnamefont {Sokolov}}, \bibinfo {author} {\bibfnamefont
  {R.}~\bibnamefont {Di~Remigio}}, \bibinfo {author} {\bibfnamefont {R.~M.}\
  \bibnamefont {Richard}}, \emph {et~al.},\ }\bibfield  {title} {\bibinfo
  {title} {Psi4 1.1: An open-source electronic structure program emphasizing
  automation, advanced libraries, and interoperability},\ }\href
  {https://doi.org/10.1021/acs.jctc.7b00174} {\bibfield  {journal} {\bibinfo
  {journal} {J. Chem. Theory Comput.}\ }\textbf {\bibinfo {volume} {13}},\
  \bibinfo {pages} {3185} (\bibinfo {year} {2017})}\BibitemShut {NoStop}%
\bibitem [{\citenamefont {Sun}\ \emph {et~al.}(2018)\citenamefont {Sun},
  \citenamefont {Berkelbach}, \citenamefont {Blunt}, \citenamefont {Booth},
  \citenamefont {Guo}, \citenamefont {Li}, \citenamefont {Liu}, \citenamefont
  {McClain}, \citenamefont {Sayfutyarova}, \citenamefont {Sharma},
  \citenamefont {Wouters},\ and\ \citenamefont {Chan}}]{sun2018pyscf}%
  \BibitemOpen
  \bibfield  {author} {\bibinfo {author} {\bibfnamefont {Q.}~\bibnamefont
  {Sun}}, \bibinfo {author} {\bibfnamefont {T.~C.}\ \bibnamefont {Berkelbach}},
  \bibinfo {author} {\bibfnamefont {N.~S.}\ \bibnamefont {Blunt}}, \bibinfo
  {author} {\bibfnamefont {G.~H.}\ \bibnamefont {Booth}}, \bibinfo {author}
  {\bibfnamefont {S.}~\bibnamefont {Guo}}, \bibinfo {author} {\bibfnamefont
  {Z.}~\bibnamefont {Li}}, \bibinfo {author} {\bibfnamefont {J.}~\bibnamefont
  {Liu}}, \bibinfo {author} {\bibfnamefont {J.~D.}\ \bibnamefont {McClain}},
  \bibinfo {author} {\bibfnamefont {E.~R.}\ \bibnamefont {Sayfutyarova}},
  \bibinfo {author} {\bibfnamefont {S.}~\bibnamefont {Sharma}}, \bibinfo
  {author} {\bibfnamefont {S.}~\bibnamefont {Wouters}},\ and\ \bibinfo {author}
  {\bibfnamefont {G.~K.-L.}\ \bibnamefont {Chan}},\ }\bibfield  {title}
  {\bibinfo {title} {Pyscf: the python-based simulations of chemistry
  framework},\ }\href {https://doi.org/10.1002/wcms.1340} {\bibfield  {journal}
  {\bibinfo  {journal} {Wiley Interdisciplinary Reviews: Computational
  Molecular Science}\ }\textbf {\bibinfo {volume} {8}},\ \bibinfo {pages}
  {e1340} (\bibinfo {year} {2018})}\BibitemShut {NoStop}%
\bibitem [{\citenamefont {Sun}(2015)}]{sun_secondary_pyscf}%
  \BibitemOpen
  \bibfield  {author} {\bibinfo {author} {\bibfnamefont {Q.}~\bibnamefont
  {Sun}},\ }\bibfield  {title} {\bibinfo {title} {Libcint: An efficient general
  integral library for gaussian basis functions},\ }\href
  {https://doi.org/10.1002/jcc.23981} {\bibfield  {journal} {\bibinfo
  {journal} {J. Comput. Chem.}\ }\textbf {\bibinfo {volume} {36}},\ \bibinfo
  {pages} {1664} (\bibinfo {year} {2015})}\BibitemShut {NoStop}%
\bibitem [{\citenamefont {Harrison}\ \emph {et~al.}(2016)\citenamefont
  {Harrison}, \citenamefont {Beylkin}, \citenamefont {Bischoff}, \citenamefont
  {Calvin}, \citenamefont {Fann}, \citenamefont {Fosso-Tande}, \citenamefont
  {Galindo}, \citenamefont {Hammond}, \citenamefont {Hartman-Baker},
  \citenamefont {Hill} \emph {et~al.}}]{harrison2016madness}%
  \BibitemOpen
  \bibfield  {author} {\bibinfo {author} {\bibfnamefont {R.~J.}\ \bibnamefont
  {Harrison}}, \bibinfo {author} {\bibfnamefont {G.}~\bibnamefont {Beylkin}},
  \bibinfo {author} {\bibfnamefont {F.~A.}\ \bibnamefont {Bischoff}}, \bibinfo
  {author} {\bibfnamefont {J.~A.}\ \bibnamefont {Calvin}}, \bibinfo {author}
  {\bibfnamefont {G.~I.}\ \bibnamefont {Fann}}, \bibinfo {author}
  {\bibfnamefont {J.}~\bibnamefont {Fosso-Tande}}, \bibinfo {author}
  {\bibfnamefont {D.}~\bibnamefont {Galindo}}, \bibinfo {author} {\bibfnamefont
  {J.~R.}\ \bibnamefont {Hammond}}, \bibinfo {author} {\bibfnamefont
  {R.}~\bibnamefont {Hartman-Baker}}, \bibinfo {author} {\bibfnamefont {J.~C.}\
  \bibnamefont {Hill}}, \emph {et~al.},\ }\bibfield  {title} {\bibinfo {title}
  {{MADNESS}: A multiresolution, adaptive numerical environment for scientific
  simulation},\ }\href {https://doi.org/10.1137/15M1026171} {\bibfield
  {journal} {\bibinfo  {journal} {SIAM J. Sci. Comput.}\ }\textbf {\bibinfo
  {volume} {38}},\ \bibinfo {pages} {S123} (\bibinfo {year}
  {2016})}\BibitemShut {NoStop}%
\bibitem [{\citenamefont {Bradbury}\ \emph {et~al.}(2018)\citenamefont
  {Bradbury}, \citenamefont {Frostig}, \citenamefont {Hawkins}, \citenamefont
  {Johnson}, \citenamefont {Leary}, \citenamefont {Maclaurin},\ and\
  \citenamefont {Wanderman-Milne}}]{jax}%
  \BibitemOpen
  \bibfield  {author} {\bibinfo {author} {\bibfnamefont {J.}~\bibnamefont
  {Bradbury}}, \bibinfo {author} {\bibfnamefont {R.}~\bibnamefont {Frostig}},
  \bibinfo {author} {\bibfnamefont {P.}~\bibnamefont {Hawkins}}, \bibinfo
  {author} {\bibfnamefont {M.~J.}\ \bibnamefont {Johnson}}, \bibinfo {author}
  {\bibfnamefont {C.}~\bibnamefont {Leary}}, \bibinfo {author} {\bibfnamefont
  {D.}~\bibnamefont {Maclaurin}},\ and\ \bibinfo {author} {\bibfnamefont
  {S.}~\bibnamefont {Wanderman-Milne}},\ }\href {http://github.com/google/jax}
  {\bibinfo {title} {{JAX}: composable transformations of {P}ython+{N}um{P}y
  programs}} (\bibinfo {year} {2018})\BibitemShut {NoStop}%
\bibitem [{\citenamefont {Kottmann}\ \emph
  {et~al.}(2020{\natexlab{a}})\citenamefont {Kottmann}, \citenamefont
  {Alperin-Lea}, \citenamefont {Tamayo-Mendoza}, \citenamefont {Alba
  Cervera-Lierta}, \citenamefont {Yen}, \citenamefont {Verteletskyi},
  \citenamefont {Schleich}, \citenamefont {Anand}, \citenamefont {Degroote},
  \citenamefont {Chaney}, \citenamefont {Kesibi}, \citenamefont {Curnow},
  \citenamefont {Izmaylov},\ and\ \citenamefont {Aspuru-Guzik}}]{tequila}%
  \BibitemOpen
  \bibfield  {author} {\bibinfo {author} {\bibfnamefont {J.~S.}\ \bibnamefont
  {Kottmann}}, \bibinfo {author} {\bibfnamefont {S.}~\bibnamefont
  {Alperin-Lea}}, \bibinfo {author} {\bibfnamefont {T.}~\bibnamefont
  {Tamayo-Mendoza}}, \bibinfo {author} {\bibfnamefont {C.~L.}\ \bibnamefont
  {Alba Cervera-Lierta}}, \bibinfo {author} {\bibfnamefont {T.-C.}\
  \bibnamefont {Yen}}, \bibinfo {author} {\bibfnamefont {V.}~\bibnamefont
  {Verteletskyi}}, \bibinfo {author} {\bibfnamefont {P.}~\bibnamefont
  {Schleich}}, \bibinfo {author} {\bibfnamefont {A.}~\bibnamefont {Anand}},
  \bibinfo {author} {\bibfnamefont {M.}~\bibnamefont {Degroote}}, \bibinfo
  {author} {\bibfnamefont {S.}~\bibnamefont {Chaney}}, \bibinfo {author}
  {\bibfnamefont {M.}~\bibnamefont {Kesibi}}, \bibinfo {author} {\bibfnamefont
  {N.~G.}\ \bibnamefont {Curnow}}, \bibinfo {author} {\bibfnamefont {A.~F.}\
  \bibnamefont {Izmaylov}},\ and\ \bibinfo {author} {\bibfnamefont
  {A.}~\bibnamefont {Aspuru-Guzik}},\ }\href
  {https://github.com/aspuru-guzik-group/tequila} {\bibinfo {title} {{Tequila}:
  A generalized development library for novel quantum algorithms}} (\bibinfo
  {year} {2020}{\natexlab{a}})\BibitemShut {NoStop}%
\bibitem [{\citenamefont {Schuld}\ \emph {et~al.}(2019)\citenamefont {Schuld},
  \citenamefont {Bergholm}, \citenamefont {Gogolin}, \citenamefont {Izaac},\
  and\ \citenamefont {Killoran}}]{schuld2019evaluating}%
  \BibitemOpen
  \bibfield  {author} {\bibinfo {author} {\bibfnamefont {M.}~\bibnamefont
  {Schuld}}, \bibinfo {author} {\bibfnamefont {V.}~\bibnamefont {Bergholm}},
  \bibinfo {author} {\bibfnamefont {C.}~\bibnamefont {Gogolin}}, \bibinfo
  {author} {\bibfnamefont {J.}~\bibnamefont {Izaac}},\ and\ \bibinfo {author}
  {\bibfnamefont {N.}~\bibnamefont {Killoran}},\ }\bibfield  {title} {\bibinfo
  {title} {Evaluating analytic gradients on quantum hardware},\ }\href
  {https://doi.org/10.1103/PhysRevA.99.032331} {\bibfield  {journal} {\bibinfo
  {journal} {Phys. Rev. A}\ }\textbf {\bibinfo {volume} {99}},\ \bibinfo
  {pages} {032331} (\bibinfo {year} {2019})}\BibitemShut {NoStop}%
\bibitem [{\citenamefont {Crooks}(2019)}]{crooks2019gradients}%
  \BibitemOpen
  \bibfield  {author} {\bibinfo {author} {\bibfnamefont {G.~E.}\ \bibnamefont
  {Crooks}},\ }\bibfield  {title} {\bibinfo {title} {Gradients of parameterized
  quantum gates using the parameter-shift rule and gate decomposition},\ }\href
  {https://arxiv.org/abs/1905.13311} {\bibfield  {journal} {\bibinfo  {journal}
  {arXiv:1905.13311 [quant-ph]}\ } (\bibinfo {year} {2019})}\BibitemShut
  {NoStop}%
\bibitem [{\citenamefont {McClean}\ \emph {et~al.}(2017)\citenamefont
  {McClean}, \citenamefont {Sung}, \citenamefont {Kivlichan}, \citenamefont
  {Cao}, \citenamefont {Dai}, \citenamefont {Fried}, \citenamefont {Gidney},
  \citenamefont {Gimby}, \citenamefont {Gokhale}, \citenamefont {Häner},
  \citenamefont {Hardikar}, \citenamefont {Havlíček}, \citenamefont
  {Higgott}, \citenamefont {Huang}, \citenamefont {Izaac}, \citenamefont
  {Jiang}, \citenamefont {Liu}, \citenamefont {McArdle}, \citenamefont
  {Neeley}, \citenamefont {O'Brien}, \citenamefont {O'Gorman}, \citenamefont
  {Ozfidan}, \citenamefont {Radin}, \citenamefont {Romero}, \citenamefont
  {Rubin}, \citenamefont {Sawaya}, \citenamefont {Setia}, \citenamefont {Sim},
  \citenamefont {Steiger}, \citenamefont {Steudtner}, \citenamefont {Sun},
  \citenamefont {Sun}, \citenamefont {Wang}, \citenamefont {Zhang},\ and\
  \citenamefont {Babbush}}]{openfermion}%
  \BibitemOpen
  \bibfield  {author} {\bibinfo {author} {\bibfnamefont {J.~R.}\ \bibnamefont
  {McClean}}, \bibinfo {author} {\bibfnamefont {K.~J.}\ \bibnamefont {Sung}},
  \bibinfo {author} {\bibfnamefont {I.~D.}\ \bibnamefont {Kivlichan}}, \bibinfo
  {author} {\bibfnamefont {Y.}~\bibnamefont {Cao}}, \bibinfo {author}
  {\bibfnamefont {C.}~\bibnamefont {Dai}}, \bibinfo {author} {\bibfnamefont
  {E.~S.}\ \bibnamefont {Fried}}, \bibinfo {author} {\bibfnamefont
  {C.}~\bibnamefont {Gidney}}, \bibinfo {author} {\bibfnamefont
  {B.}~\bibnamefont {Gimby}}, \bibinfo {author} {\bibfnamefont
  {P.}~\bibnamefont {Gokhale}}, \bibinfo {author} {\bibfnamefont
  {T.}~\bibnamefont {Häner}}, \bibinfo {author} {\bibfnamefont
  {T.}~\bibnamefont {Hardikar}}, \bibinfo {author} {\bibfnamefont
  {V.}~\bibnamefont {Havlíček}}, \bibinfo {author} {\bibfnamefont
  {O.}~\bibnamefont {Higgott}}, \bibinfo {author} {\bibfnamefont
  {C.}~\bibnamefont {Huang}}, \bibinfo {author} {\bibfnamefont
  {J.}~\bibnamefont {Izaac}}, \bibinfo {author} {\bibfnamefont
  {Z.}~\bibnamefont {Jiang}}, \bibinfo {author} {\bibfnamefont
  {X.}~\bibnamefont {Liu}}, \bibinfo {author} {\bibfnamefont {S.}~\bibnamefont
  {McArdle}}, \bibinfo {author} {\bibfnamefont {M.}~\bibnamefont {Neeley}},
  \bibinfo {author} {\bibfnamefont {T.}~\bibnamefont {O'Brien}}, \bibinfo
  {author} {\bibfnamefont {B.}~\bibnamefont {O'Gorman}}, \bibinfo {author}
  {\bibfnamefont {I.}~\bibnamefont {Ozfidan}}, \bibinfo {author} {\bibfnamefont
  {M.~D.}\ \bibnamefont {Radin}}, \bibinfo {author} {\bibfnamefont
  {J.}~\bibnamefont {Romero}}, \bibinfo {author} {\bibfnamefont
  {N.}~\bibnamefont {Rubin}}, \bibinfo {author} {\bibfnamefont {N.~P.~D.}\
  \bibnamefont {Sawaya}}, \bibinfo {author} {\bibfnamefont {K.}~\bibnamefont
  {Setia}}, \bibinfo {author} {\bibfnamefont {S.}~\bibnamefont {Sim}}, \bibinfo
  {author} {\bibfnamefont {D.~S.}\ \bibnamefont {Steiger}}, \bibinfo {author}
  {\bibfnamefont {M.}~\bibnamefont {Steudtner}}, \bibinfo {author}
  {\bibfnamefont {Q.}~\bibnamefont {Sun}}, \bibinfo {author} {\bibfnamefont
  {W.}~\bibnamefont {Sun}}, \bibinfo {author} {\bibfnamefont {D.}~\bibnamefont
  {Wang}}, \bibinfo {author} {\bibfnamefont {F.}~\bibnamefont {Zhang}},\ and\
  \bibinfo {author} {\bibfnamefont {R.}~\bibnamefont {Babbush}},\ }\bibfield
  {title} {\bibinfo {title} {Openfermion: The electronic structure package for
  quantum computers},\ }\href {https://arxiv.org/abs/1710.07629} {\bibfield
  {journal} {\bibinfo  {journal} {arXiv:1710.07629 [quant-ph]}\ } (\bibinfo
  {year} {2017})}\BibitemShut {NoStop}%
\bibitem [{\citenamefont {Sawaya}\ \emph {et~al.}(2020)\citenamefont {Sawaya},
  \citenamefont {Menke}, \citenamefont {Kyaw}, \citenamefont {Johri},
  \citenamefont {Aspuru-Guzik},\ and\ \citenamefont
  {Guerreschi}}]{nicolas2019}%
  \BibitemOpen
  \bibfield  {author} {\bibinfo {author} {\bibfnamefont {N.~P.~D.}\
  \bibnamefont {Sawaya}}, \bibinfo {author} {\bibfnamefont {T.}~\bibnamefont
  {Menke}}, \bibinfo {author} {\bibfnamefont {T.~H.}\ \bibnamefont {Kyaw}},
  \bibinfo {author} {\bibfnamefont {S.}~\bibnamefont {Johri}}, \bibinfo
  {author} {\bibfnamefont {A.}~\bibnamefont {Aspuru-Guzik}},\ and\ \bibinfo
  {author} {\bibfnamefont {G.~G.}\ \bibnamefont {Guerreschi}},\ }\bibfield
  {title} {\bibinfo {title} {{Resource-efficient digital quantum simulation of
  $ d $-level systems for photonic, vibrational, and spin-$ s $
  Hamiltonians}},\ }\href {https://doi.org/10.1038/s41534-020-0278-0}
  {\bibfield  {journal} {\bibinfo  {journal} {npj Quantum Inf.}\ }\textbf
  {\bibinfo {volume} {6}},\ \bibinfo {pages} {49} (\bibinfo {year}
  {2020})}\BibitemShut {NoStop}%
\bibitem [{\citenamefont {Yen}\ \emph {et~al.}(2020)\citenamefont {Yen},
  \citenamefont {Verteletskyi},\ and\ \citenamefont {Izmaylov}}]{yen2020}%
  \BibitemOpen
  \bibfield  {author} {\bibinfo {author} {\bibfnamefont {T.-C.}\ \bibnamefont
  {Yen}}, \bibinfo {author} {\bibfnamefont {V.}~\bibnamefont {Verteletskyi}},\
  and\ \bibinfo {author} {\bibfnamefont {A.~F.}\ \bibnamefont {Izmaylov}},\
  }\bibfield  {title} {\bibinfo {title} {Measuring all compatible operators in
  one series of single-qubit measurements using unitary transformations},\
  }\href {https://doi.org/10.1021/acs.jctc.0c00008} {\bibfield  {journal}
  {\bibinfo  {journal} {J. Chem. Theory Comput.}\ }\textbf {\bibinfo {volume}
  {16}},\ \bibinfo {pages} {2400} (\bibinfo {year} {2020})}\BibitemShut
  {NoStop}%
\bibitem [{\citenamefont {Verteletskyi}\ \emph {et~al.}(2020)\citenamefont
  {Verteletskyi}, \citenamefont {Yen},\ and\ \citenamefont
  {Izmaylov}}]{verteletskyi2020}%
  \BibitemOpen
  \bibfield  {author} {\bibinfo {author} {\bibfnamefont {V.}~\bibnamefont
  {Verteletskyi}}, \bibinfo {author} {\bibfnamefont {T.-C.}\ \bibnamefont
  {Yen}},\ and\ \bibinfo {author} {\bibfnamefont {A.~F.}\ \bibnamefont
  {Izmaylov}},\ }\bibfield  {title} {\bibinfo {title} {Measurement optimization
  in the variational quantum eigensolver using a minimum clique cover},\ }\href
  {https://doi.org/10.1063/1.5141458} {\bibfield  {journal} {\bibinfo
  {journal} {J. Chem. Phys.}\ }\textbf {\bibinfo {volume} {152}},\ \bibinfo
  {pages} {124114} (\bibinfo {year} {2020})}\BibitemShut {NoStop}%
\bibitem [{\citenamefont {Sivarajah}\ \emph {et~al.}(2020)\citenamefont
  {Sivarajah}, \citenamefont {Dilkes}, \citenamefont {Cowtan}, \citenamefont
  {Simmons}, \citenamefont {Edgington},\ and\ \citenamefont {Duncan}}]{tket}%
  \BibitemOpen
  \bibfield  {author} {\bibinfo {author} {\bibfnamefont {S.}~\bibnamefont
  {Sivarajah}}, \bibinfo {author} {\bibfnamefont {S.}~\bibnamefont {Dilkes}},
  \bibinfo {author} {\bibfnamefont {A.}~\bibnamefont {Cowtan}}, \bibinfo
  {author} {\bibfnamefont {W.}~\bibnamefont {Simmons}}, \bibinfo {author}
  {\bibfnamefont {A.}~\bibnamefont {Edgington}},\ and\ \bibinfo {author}
  {\bibfnamefont {R.}~\bibnamefont {Duncan}},\ }\bibfield  {title} {\bibinfo
  {title} {{t$|ket\rangle$: A retargetable compiler for NISQ devices}},\ }\href
  {http://iopscience.iop.org/10.1088/2058-9565/ab8e92} {\bibfield  {journal}
  {\bibinfo  {journal} {Quantum Sci. Technol.}\ } (\bibinfo {year}
  {2020})}\BibitemShut {NoStop}%
\bibitem [{\citenamefont {Kissinger}\ and\ \citenamefont {van~de
  Wetering}(2020)}]{kissinger2020Pyzx}%
  \BibitemOpen
  \bibfield  {author} {\bibinfo {author} {\bibfnamefont {A.}~\bibnamefont
  {Kissinger}}\ and\ \bibinfo {author} {\bibfnamefont {J.}~\bibnamefont {van~de
  Wetering}},\ }\bibfield  {title} {\bibinfo {title} {{PyZX: Large Scale
  Automated Diagrammatic Reasoning}},\ }in\ \href
  {https://doi.org/10.4204/EPTCS.318.14} {\emph {\bibinfo {booktitle} {{\rm
  Proceedings 16th International Conference on} Quantum Physics and Logic, {\rm
  Chapman University, Orange, CA, USA., 10-14 June 2019}}}},\ \bibinfo {series}
  {Electronic Proceedings in Theoretical Computer Science}, Vol.\ \bibinfo
  {volume} {318},\ \bibinfo {editor} {edited by\ \bibinfo {editor}
  {\bibfnamefont {B.}~\bibnamefont {Coecke}}\ and\ \bibinfo {editor}
  {\bibfnamefont {M.}~\bibnamefont {Leifer}}}\ (\bibinfo  {publisher} {Open
  Publishing Association},\ \bibinfo {year} {2020})\ pp.\ \bibinfo {pages}
  {229--241}\BibitemShut {NoStop}%
\bibitem [{\citenamefont {{Virtanen}}\ \emph {et~al.}(2020)\citenamefont
  {{Virtanen}}, \citenamefont {{Gommers}}, \citenamefont {{Oliphant}},
  \citenamefont {{Haberland}}, \citenamefont {{Reddy}}, \citenamefont
  {{Cournapeau}}, \citenamefont {{Burovski}}, \citenamefont {{Peterson}},
  \citenamefont {{Weckesser}}, \citenamefont {{Bright}}, \citenamefont {{van
  der Walt}}, \citenamefont {{Brett}}, \citenamefont {{Wilson}}, \citenamefont
  {{Jarrod Millman}}, \citenamefont {{Mayorov}}, \citenamefont {{Nelson}},
  \citenamefont {{Jones}}, \citenamefont {{Kern}}, \citenamefont {{Larson}},
  \citenamefont {{Carey}}, \citenamefont {{Polat}}, \citenamefont {{Feng}},
  \citenamefont {{Moore}}, \citenamefont {{Vand erPlas}}, \citenamefont
  {{Laxalde}}, \citenamefont {{Perktold}}, \citenamefont {{Cimrman}},
  \citenamefont {{Henriksen}}, \citenamefont {{Quintero}}, \citenamefont
  {{Harris}}, \citenamefont {{Archibald}}, \citenamefont {{Ribeiro}},
  \citenamefont {{Pedregosa}}, \citenamefont {{van Mulbregt}},\ and\
  \citenamefont {{Contributors}}}]{scipy}%
  \BibitemOpen
  \bibfield  {author} {\bibinfo {author} {\bibfnamefont {P.}~\bibnamefont
  {{Virtanen}}}, \bibinfo {author} {\bibfnamefont {R.}~\bibnamefont
  {{Gommers}}}, \bibinfo {author} {\bibfnamefont {T.~E.}\ \bibnamefont
  {{Oliphant}}}, \bibinfo {author} {\bibfnamefont {M.}~\bibnamefont
  {{Haberland}}}, \bibinfo {author} {\bibfnamefont {T.}~\bibnamefont
  {{Reddy}}}, \bibinfo {author} {\bibfnamefont {D.}~\bibnamefont
  {{Cournapeau}}}, \bibinfo {author} {\bibfnamefont {E.}~\bibnamefont
  {{Burovski}}}, \bibinfo {author} {\bibfnamefont {P.}~\bibnamefont
  {{Peterson}}}, \bibinfo {author} {\bibfnamefont {W.}~\bibnamefont
  {{Weckesser}}}, \bibinfo {author} {\bibfnamefont {J.}~\bibnamefont
  {{Bright}}}, \bibinfo {author} {\bibfnamefont {S.~J.}\ \bibnamefont {{van der
  Walt}}}, \bibinfo {author} {\bibfnamefont {M.}~\bibnamefont {{Brett}}},
  \bibinfo {author} {\bibfnamefont {J.}~\bibnamefont {{Wilson}}}, \bibinfo
  {author} {\bibfnamefont {K.}~\bibnamefont {{Jarrod Millman}}}, \bibinfo
  {author} {\bibfnamefont {N.}~\bibnamefont {{Mayorov}}}, \bibinfo {author}
  {\bibfnamefont {A.~R.~J.}\ \bibnamefont {{Nelson}}}, \bibinfo {author}
  {\bibfnamefont {E.}~\bibnamefont {{Jones}}}, \bibinfo {author} {\bibfnamefont
  {R.}~\bibnamefont {{Kern}}}, \bibinfo {author} {\bibfnamefont
  {E.}~\bibnamefont {{Larson}}}, \bibinfo {author} {\bibfnamefont
  {C.}~\bibnamefont {{Carey}}}, \bibinfo {author} {\bibfnamefont
  {{\.I}.}~\bibnamefont {{Polat}}}, \bibinfo {author} {\bibfnamefont
  {Y.}~\bibnamefont {{Feng}}}, \bibinfo {author} {\bibfnamefont {E.~W.}\
  \bibnamefont {{Moore}}}, \bibinfo {author} {\bibfnamefont {J.}~\bibnamefont
  {{Vand erPlas}}}, \bibinfo {author} {\bibfnamefont {D.}~\bibnamefont
  {{Laxalde}}}, \bibinfo {author} {\bibfnamefont {J.}~\bibnamefont
  {{Perktold}}}, \bibinfo {author} {\bibfnamefont {R.}~\bibnamefont
  {{Cimrman}}}, \bibinfo {author} {\bibfnamefont {I.}~\bibnamefont
  {{Henriksen}}}, \bibinfo {author} {\bibfnamefont {E.~A.}\ \bibnamefont
  {{Quintero}}}, \bibinfo {author} {\bibfnamefont {C.~R.}\ \bibnamefont
  {{Harris}}}, \bibinfo {author} {\bibfnamefont {A.~M.}\ \bibnamefont
  {{Archibald}}}, \bibinfo {author} {\bibfnamefont {A.~H.}\ \bibnamefont
  {{Ribeiro}}}, \bibinfo {author} {\bibfnamefont {F.}~\bibnamefont
  {{Pedregosa}}}, \bibinfo {author} {\bibfnamefont {P.}~\bibnamefont {{van
  Mulbregt}}},\ and\ \bibinfo {author} {\bibfnamefont {S.~.~.}\ \bibnamefont
  {{Contributors}}},\ }\bibfield  {title} {\bibinfo {title} {{SciPy 1.0:
  Fundamental Algorithms for Scientific Computing in Python}},\ }\href
  {https://doi.org/10.1038/s41592-019-0686-2} {\bibfield  {journal} {\bibinfo
  {journal} {Nature Methods}\ }\textbf {\bibinfo {volume} {17}},\ \bibinfo
  {pages} {261} (\bibinfo {year} {2020})}\BibitemShut {NoStop}%
\bibitem [{\citenamefont {authors}(2016)}]{gpyopt2016}%
  \BibitemOpen
  \bibfield  {author} {\bibinfo {author} {\bibfnamefont {T.~G.}\ \bibnamefont
  {authors}},\ }\href {http://github.com/SheffieldML/GPyOpt} {\bibinfo {title}
  {{GPyOpt}: A bayesian optimization framework in python}} (\bibinfo {year}
  {2016})\BibitemShut {NoStop}%
\bibitem [{\citenamefont {Häse}\ \emph {et~al.}(2018)\citenamefont {Häse},
  \citenamefont {Roch}, \citenamefont {Kreisbeck},\ and\ \citenamefont
  {Aspuru-Guzik}}]{hase2018phoenics}%
  \BibitemOpen
  \bibfield  {author} {\bibinfo {author} {\bibfnamefont {F.}~\bibnamefont
  {Häse}}, \bibinfo {author} {\bibfnamefont {L.~M.}\ \bibnamefont {Roch}},
  \bibinfo {author} {\bibfnamefont {C.}~\bibnamefont {Kreisbeck}},\ and\
  \bibinfo {author} {\bibfnamefont {A.}~\bibnamefont {Aspuru-Guzik}},\
  }\bibfield  {title} {\bibinfo {title} {Phoenics: A bayesian optimizer for
  chemistry},\ }\href {https://doi.org/10.1021/acscentsci.8b00307} {\bibfield
  {journal} {\bibinfo  {journal} {ACS Cent. Sci.}\ }\textbf {\bibinfo {volume}
  {4}},\ \bibinfo {pages} {1134} (\bibinfo {year} {2018})}\BibitemShut
  {NoStop}%
\bibitem [{\citenamefont {Stokes}\ \emph {et~al.}(2020)\citenamefont {Stokes},
  \citenamefont {Izaac}, \citenamefont {Killoran},\ and\ \citenamefont
  {Carleo}}]{Stokes2020quantumnatural}%
  \BibitemOpen
  \bibfield  {author} {\bibinfo {author} {\bibfnamefont {J.}~\bibnamefont
  {Stokes}}, \bibinfo {author} {\bibfnamefont {J.}~\bibnamefont {Izaac}},
  \bibinfo {author} {\bibfnamefont {N.}~\bibnamefont {Killoran}},\ and\
  \bibinfo {author} {\bibfnamefont {G.}~\bibnamefont {Carleo}},\ }\bibfield
  {title} {\bibinfo {title} {Quantum {N}atural {G}radient},\ }\href
  {https://doi.org/10.22331/q-2020-05-25-269} {\bibfield  {journal} {\bibinfo
  {journal} {{Quantum}}\ }\textbf {\bibinfo {volume} {4}},\ \bibinfo {pages}
  {269} (\bibinfo {year} {2020})}\BibitemShut {NoStop}%
\bibitem [{\citenamefont {Zhu}\ \emph {et~al.}(2019)\citenamefont {Zhu},
  \citenamefont {Linke}, \citenamefont {Benedetti}, \citenamefont {Landsman},
  \citenamefont {Nguyen}, \citenamefont {Alderete}, \citenamefont
  {Perdomo-Ortiz}, \citenamefont {Korda}, \citenamefont {Garfoot},
  \citenamefont {Brecque}, \citenamefont {Egan}, \citenamefont {Perdomo},\ and\
  \citenamefont {Monroe}}]{zhu2019classifier}%
  \BibitemOpen
  \bibfield  {author} {\bibinfo {author} {\bibfnamefont {D.}~\bibnamefont
  {Zhu}}, \bibinfo {author} {\bibfnamefont {N.~M.}\ \bibnamefont {Linke}},
  \bibinfo {author} {\bibfnamefont {M.}~\bibnamefont {Benedetti}}, \bibinfo
  {author} {\bibfnamefont {K.~A.}\ \bibnamefont {Landsman}}, \bibinfo {author}
  {\bibfnamefont {N.~H.}\ \bibnamefont {Nguyen}}, \bibinfo {author}
  {\bibfnamefont {C.~H.}\ \bibnamefont {Alderete}}, \bibinfo {author}
  {\bibfnamefont {A.}~\bibnamefont {Perdomo-Ortiz}}, \bibinfo {author}
  {\bibfnamefont {N.}~\bibnamefont {Korda}}, \bibinfo {author} {\bibfnamefont
  {A.}~\bibnamefont {Garfoot}}, \bibinfo {author} {\bibfnamefont
  {C.}~\bibnamefont {Brecque}}, \bibinfo {author} {\bibfnamefont
  {L.}~\bibnamefont {Egan}}, \bibinfo {author} {\bibfnamefont {O.}~\bibnamefont
  {Perdomo}},\ and\ \bibinfo {author} {\bibfnamefont {C.}~\bibnamefont
  {Monroe}},\ }\bibfield  {title} {\bibinfo {title} {Training of quantum
  circuits on a hybrid quantum computer},\ }\bibfield  {journal} {\bibinfo
  {journal} {Sci. Adv.}\ }\textbf {\bibinfo {volume} {5}},\ \href
  {https://doi.org/10.1126/sciadv.aaw9918} {10.1126/sciadv.aaw9918} (\bibinfo
  {year} {2019})\BibitemShut {NoStop}%
\bibitem [{\citenamefont {Kottmann}\ \emph
  {et~al.}(2021{\natexlab{a}})\citenamefont {Kottmann}, \citenamefont
  {Schleich}, \citenamefont {Tamayo-Mendoza},\ and\ \citenamefont
  {Aspuru-Guzik}}]{kottmann2020reducing}%
  \BibitemOpen
  \bibfield  {author} {\bibinfo {author} {\bibfnamefont {J.~S.}\ \bibnamefont
  {Kottmann}}, \bibinfo {author} {\bibfnamefont {P.}~\bibnamefont {Schleich}},
  \bibinfo {author} {\bibfnamefont {T.}~\bibnamefont {Tamayo-Mendoza}},\ and\
  \bibinfo {author} {\bibfnamefont {A.}~\bibnamefont {Aspuru-Guzik}},\
  }\bibfield  {title} {\bibinfo {title} {Reducing qubit requirements while
  maintaining numerical precision for the variational quantum eigensolver: A
  basis-set-free approach},\ }\href
  {https://doi.org/10.1021/acs.jpclett.0c03410} {\bibfield  {journal} {\bibinfo
   {journal} {J. Phys. Chem. Lett.}\ }\textbf {\bibinfo {volume} {12}},\
  \bibinfo {pages} {663} (\bibinfo {year} {2021}{\natexlab{a}})}\BibitemShut
  {NoStop}%
\bibitem [{\citenamefont {Kottmann}\ \emph
  {et~al.}(2020{\natexlab{b}})\citenamefont {Kottmann}, \citenamefont {Krenn},
  \citenamefont {Kyaw}, \citenamefont {Alperin-Lea},\ and\ \citenamefont
  {Aspuru-Guzik}}]{kottmann2020quantum}%
  \BibitemOpen
  \bibfield  {author} {\bibinfo {author} {\bibfnamefont {J.~S.}\ \bibnamefont
  {Kottmann}}, \bibinfo {author} {\bibfnamefont {M.}~\bibnamefont {Krenn}},
  \bibinfo {author} {\bibfnamefont {T.~H.}\ \bibnamefont {Kyaw}}, \bibinfo
  {author} {\bibfnamefont {S.}~\bibnamefont {Alperin-Lea}},\ and\ \bibinfo
  {author} {\bibfnamefont {A.}~\bibnamefont {Aspuru-Guzik}},\ }\bibfield
  {title} {\bibinfo {title} {Quantum computer-aided design of quantum optics
  hardware},\ }\href {https://arxiv.org/abs/2006.03075} {\bibfield  {journal}
  {\bibinfo  {journal} {arXiv:2006.03075 [quant-ph]}\ } (\bibinfo {year}
  {2020}{\natexlab{b}})}\BibitemShut {NoStop}%
\bibitem [{\citenamefont {Cervera-Lierta}\ \emph {et~al.}(2020)\citenamefont
  {Cervera-Lierta}, \citenamefont {Kottmann},\ and\ \citenamefont
  {Aspuru-Guzik}}]{cerveralierta2020metavariational}%
  \BibitemOpen
  \bibfield  {author} {\bibinfo {author} {\bibfnamefont {A.}~\bibnamefont
  {Cervera-Lierta}}, \bibinfo {author} {\bibfnamefont {J.~S.}\ \bibnamefont
  {Kottmann}},\ and\ \bibinfo {author} {\bibfnamefont {A.}~\bibnamefont
  {Aspuru-Guzik}},\ }\bibfield  {title} {\bibinfo {title} {The meta-variational
  quantum eigensolver (meta-vqe): Learning energy profiles of parameterized
  hamiltonians for quantum simulation},\ }\href
  {https://arxiv.org/abs/2009.13545} {\bibfield  {journal} {\bibinfo  {journal}
  {arXiv:2009.13545[quant-ph]}\ } (\bibinfo {year} {2020})}\BibitemShut
  {NoStop}%
\bibitem [{\citenamefont {Reiher}\ \emph {et~al.}(2017)\citenamefont {Reiher},
  \citenamefont {Wiebe}, \citenamefont {Svore}, \citenamefont {Wecker},\ and\
  \citenamefont {Troyer}}]{reiher2017elucidating}%
  \BibitemOpen
  \bibfield  {author} {\bibinfo {author} {\bibfnamefont {M.}~\bibnamefont
  {Reiher}}, \bibinfo {author} {\bibfnamefont {N.}~\bibnamefont {Wiebe}},
  \bibinfo {author} {\bibfnamefont {K.~M.}\ \bibnamefont {Svore}}, \bibinfo
  {author} {\bibfnamefont {D.}~\bibnamefont {Wecker}},\ and\ \bibinfo {author}
  {\bibfnamefont {M.}~\bibnamefont {Troyer}},\ }\bibfield  {title} {\bibinfo
  {title} {Elucidating reaction mechanisms on quantum computers},\ }\href
  {https://doi.org/10.1073/pnas.1619152114} {\bibfield  {journal} {\bibinfo
  {journal} {PNAS}\ }\textbf {\bibinfo {volume} {114}},\ \bibinfo {pages}
  {7555} (\bibinfo {year} {2017})}\BibitemShut {NoStop}%
\bibitem [{\citenamefont {McClean}\ \emph {et~al.}(2016)\citenamefont
  {McClean}, \citenamefont {Romero}, \citenamefont {Babbush},\ and\
  \citenamefont {Aspuru-Guzik}}]{McClean2016theoryofvqe}%
  \BibitemOpen
  \bibfield  {author} {\bibinfo {author} {\bibfnamefont {J.~R.}\ \bibnamefont
  {McClean}}, \bibinfo {author} {\bibfnamefont {J.}~\bibnamefont {Romero}},
  \bibinfo {author} {\bibfnamefont {R.}~\bibnamefont {Babbush}},\ and\ \bibinfo
  {author} {\bibfnamefont {A.}~\bibnamefont {Aspuru-Guzik}},\ }\bibfield
  {title} {\bibinfo {title} {The theory of variational hybrid quantum-classical
  algorithms},\ }\href {https://doi.org/10.1088/1367-2630/18/2/023023}
  {\bibfield  {journal} {\bibinfo  {journal} {New J. Phys.}\ }\textbf {\bibinfo
  {volume} {18}},\ \bibinfo {pages} {023023} (\bibinfo {year}
  {2016})}\BibitemShut {NoStop}%
\bibitem [{\citenamefont {Helgaker}\ \emph {et~al.}(2014)\citenamefont
  {Helgaker}, \citenamefont {Jorgensen},\ and\ \citenamefont
  {Olsen}}]{helgaker2014molecular}%
  \BibitemOpen
  \bibfield  {author} {\bibinfo {author} {\bibfnamefont {T.}~\bibnamefont
  {Helgaker}}, \bibinfo {author} {\bibfnamefont {P.}~\bibnamefont
  {Jorgensen}},\ and\ \bibinfo {author} {\bibfnamefont {J.}~\bibnamefont
  {Olsen}},\ }\href {https://doi.org/10.1002/9781119019572} {\emph {\bibinfo
  {title} {Molecular electronic-structure theory}}}\ (\bibinfo  {publisher}
  {John Wiley \& Sons},\ \bibinfo {year} {2014})\BibitemShut {NoStop}%
\bibitem [{\citenamefont {Shavitt}\ and\ \citenamefont
  {Bartlett}(2009)}]{shavitt2009many}%
  \BibitemOpen
  \bibfield  {author} {\bibinfo {author} {\bibfnamefont {I.}~\bibnamefont
  {Shavitt}}\ and\ \bibinfo {author} {\bibfnamefont {R.~J.}\ \bibnamefont
  {Bartlett}},\ }\href {https://doi.org/10.1017/CBO9780511596834} {\emph
  {\bibinfo {title} {Many-body methods in chemistry and physics: {MBPT} and
  coupled-cluster theory}}}\ (\bibinfo  {publisher} {Cambridge university
  press},\ \bibinfo {year} {2009})\BibitemShut {NoStop}%
\bibitem [{\citenamefont {J{\o}rgensen}(2012)}]{jorgensen2012second}%
  \BibitemOpen
  \bibfield  {author} {\bibinfo {author} {\bibfnamefont {P.}~\bibnamefont
  {J{\o}rgensen}},\ }\href
  {https://www.sciencedirect.com/book/9780123902207/second-quantization-based-methods-in-quantum-chemistry}
  {\emph {\bibinfo {title} {Second quantization-based methods in quantum
  chemistry}}}\ (\bibinfo  {publisher} {Elsevier},\ \bibinfo {year}
  {2012})\BibitemShut {NoStop}%
\bibitem [{\citenamefont {Surj{\'a}n}(2012)}]{surjan2012second}%
  \BibitemOpen
  \bibfield  {author} {\bibinfo {author} {\bibfnamefont {P.~R.}\ \bibnamefont
  {Surj{\'a}n}},\ }\href
  {https://www.springer.com/gp/book/9783642747571#aboutBook} {\emph {\bibinfo
  {title} {Second quantized approach to quantum chemistry: an elementary
  introduction}}}\ (\bibinfo  {publisher} {Springer Science \& Business
  Media},\ \bibinfo {year} {2012})\BibitemShut {NoStop}%
\bibitem [{\citenamefont {Cao}\ \emph {et~al.}(2019)\citenamefont {Cao},
  \citenamefont {Romero}, \citenamefont {Olson}, \citenamefont {Degroote},
  \citenamefont {Johnson}, \citenamefont {Kieferov{\'a}}, \citenamefont
  {Kivlichan}, \citenamefont {Menke}, \citenamefont {Peropadre}, \citenamefont
  {Sawaya} \emph {et~al.}}]{cao2019quantumreview}%
  \BibitemOpen
  \bibfield  {author} {\bibinfo {author} {\bibfnamefont {Y.}~\bibnamefont
  {Cao}}, \bibinfo {author} {\bibfnamefont {J.}~\bibnamefont {Romero}},
  \bibinfo {author} {\bibfnamefont {J.~P.}\ \bibnamefont {Olson}}, \bibinfo
  {author} {\bibfnamefont {M.}~\bibnamefont {Degroote}}, \bibinfo {author}
  {\bibfnamefont {P.~D.}\ \bibnamefont {Johnson}}, \bibinfo {author}
  {\bibfnamefont {M.}~\bibnamefont {Kieferov{\'a}}}, \bibinfo {author}
  {\bibfnamefont {I.~D.}\ \bibnamefont {Kivlichan}}, \bibinfo {author}
  {\bibfnamefont {T.}~\bibnamefont {Menke}}, \bibinfo {author} {\bibfnamefont
  {B.}~\bibnamefont {Peropadre}}, \bibinfo {author} {\bibfnamefont {N.~P.}\
  \bibnamefont {Sawaya}}, \emph {et~al.},\ }\bibfield  {title} {\bibinfo
  {title} {Quantum chemistry in the age of quantum computing},\ }\href
  {https://doi.org/10.1021/acs.chemrev.8b00803} {\bibfield  {journal} {\bibinfo
   {journal} {Chem. Rev.}\ }\textbf {\bibinfo {volume} {119}},\ \bibinfo
  {pages} {10856} (\bibinfo {year} {2019})}\BibitemShut {NoStop}%
\bibitem [{\citenamefont {McArdle}\ \emph {et~al.}(2020)\citenamefont
  {McArdle}, \citenamefont {Endo}, \citenamefont {Aspuru-Guzik}, \citenamefont
  {Benjamin},\ and\ \citenamefont {Yuan}}]{mcardle2020quantumreview}%
  \BibitemOpen
  \bibfield  {author} {\bibinfo {author} {\bibfnamefont {S.}~\bibnamefont
  {McArdle}}, \bibinfo {author} {\bibfnamefont {S.}~\bibnamefont {Endo}},
  \bibinfo {author} {\bibfnamefont {A.}~\bibnamefont {Aspuru-Guzik}}, \bibinfo
  {author} {\bibfnamefont {S.~C.}\ \bibnamefont {Benjamin}},\ and\ \bibinfo
  {author} {\bibfnamefont {X.}~\bibnamefont {Yuan}},\ }\bibfield  {title}
  {\bibinfo {title} {Quantum computational chemistry},\ }\href
  {https://doi.org/10.1103/RevModPhys.92.015003} {\bibfield  {journal}
  {\bibinfo  {journal} {Rev. Mod. Phys.}\ }\textbf {\bibinfo {volume} {92}},\
  \bibinfo {pages} {015003} (\bibinfo {year} {2020})}\BibitemShut {NoStop}%
\bibitem [{\citenamefont {Fermann}\ and\ \citenamefont
  {Valeev}(2020)}]{fermann2020fundamentals}%
  \BibitemOpen
  \bibfield  {author} {\bibinfo {author} {\bibfnamefont {J.~T.}\ \bibnamefont
  {Fermann}}\ and\ \bibinfo {author} {\bibfnamefont {E.~F.}\ \bibnamefont
  {Valeev}},\ }\bibfield  {title} {\bibinfo {title} {Fundamentals of molecular
  integrals evaluation},\ }\href {https://arxiv.org/abs/2007.12057} {\bibfield
  {journal} {\bibinfo  {journal} {arXiv:2007.12057 [quant-ph]}\ } (\bibinfo
  {year} {2020})}\BibitemShut {NoStop}%
\bibitem [{\citenamefont {Bravyi}\ and\ \citenamefont
  {Kitaev}(2002)}]{bravyi2002}%
  \BibitemOpen
  \bibfield  {author} {\bibinfo {author} {\bibfnamefont {S.~B.}\ \bibnamefont
  {Bravyi}}\ and\ \bibinfo {author} {\bibfnamefont {A.~Y.}\ \bibnamefont
  {Kitaev}},\ }\bibfield  {title} {\bibinfo {title} {Fermionic quantum
  computation},\ }\href
  {https://doi.org/https://doi.org/10.1006/aphy.2002.6254} {\bibfield
  {journal} {\bibinfo  {journal} {Ann. of Physics}\ }\textbf {\bibinfo {volume}
  {298}},\ \bibinfo {pages} {210 } (\bibinfo {year} {2002})}\BibitemShut
  {NoStop}%
\bibitem [{\citenamefont {Seeley}\ \emph {et~al.}(2012)\citenamefont {Seeley},
  \citenamefont {Richard},\ and\ \citenamefont {Love}}]{seeley2012}%
  \BibitemOpen
  \bibfield  {author} {\bibinfo {author} {\bibfnamefont {J.~T.}\ \bibnamefont
  {Seeley}}, \bibinfo {author} {\bibfnamefont {M.~J.}\ \bibnamefont
  {Richard}},\ and\ \bibinfo {author} {\bibfnamefont {P.~J.}\ \bibnamefont
  {Love}},\ }\bibfield  {title} {\bibinfo {title} {The {Bravyi-Kitaev}
  transformation for quantum computation of electronic structure},\ }\href
  {https://doi.org/10.1063/1.4768229} {\bibfield  {journal} {\bibinfo
  {journal} {J. Chem. Phys.}\ }\textbf {\bibinfo {volume} {137}},\ \bibinfo
  {pages} {224109} (\bibinfo {year} {2012})}\BibitemShut {NoStop}%
\bibitem [{\citenamefont {Setia}\ and\ \citenamefont
  {Whitfield}(2018)}]{setia2018}%
  \BibitemOpen
  \bibfield  {author} {\bibinfo {author} {\bibfnamefont {K.}~\bibnamefont
  {Setia}}\ and\ \bibinfo {author} {\bibfnamefont {J.~D.}\ \bibnamefont
  {Whitfield}},\ }\bibfield  {title} {\bibinfo {title} {{Bravyi-Kitaev}
  superfast simulation of electronic structure on a quantum computer},\ }\href
  {https://doi.org/10.1063/1.5019371} {\bibfield  {journal} {\bibinfo
  {journal} {J. Chem. Phys.}\ }\textbf {\bibinfo {volume} {148}},\ \bibinfo
  {pages} {164104} (\bibinfo {year} {2018})}\BibitemShut {NoStop}%
\bibitem [{\citenamefont {Smith}\ \emph {et~al.}(2020)\citenamefont {Smith},
  \citenamefont {Burns}, \citenamefont {Simmonett}, \citenamefont {Parrish},
  \citenamefont {Schieber}, \citenamefont {Galvelis}, \citenamefont {Kraus},
  \citenamefont {Kruse}, \citenamefont {Di~Remigio}, \citenamefont {Alenaizan}
  \emph {et~al.}}]{psi42020}%
  \BibitemOpen
  \bibfield  {author} {\bibinfo {author} {\bibfnamefont {D.~G.}\ \bibnamefont
  {Smith}}, \bibinfo {author} {\bibfnamefont {L.~A.}\ \bibnamefont {Burns}},
  \bibinfo {author} {\bibfnamefont {A.~C.}\ \bibnamefont {Simmonett}}, \bibinfo
  {author} {\bibfnamefont {R.~M.}\ \bibnamefont {Parrish}}, \bibinfo {author}
  {\bibfnamefont {M.~C.}\ \bibnamefont {Schieber}}, \bibinfo {author}
  {\bibfnamefont {R.}~\bibnamefont {Galvelis}}, \bibinfo {author}
  {\bibfnamefont {P.}~\bibnamefont {Kraus}}, \bibinfo {author} {\bibfnamefont
  {H.}~\bibnamefont {Kruse}}, \bibinfo {author} {\bibfnamefont
  {R.}~\bibnamefont {Di~Remigio}}, \bibinfo {author} {\bibfnamefont
  {A.}~\bibnamefont {Alenaizan}}, \emph {et~al.},\ }\bibfield  {title}
  {\bibinfo {title} {{PSI4} 1.4: Open-source software for high-throughput
  quantum chemistry},\ }\href {https://doi.org/10.1063/5.0006002} {\bibfield
  {journal} {\bibinfo  {journal} {J. Chem. Phys.}\ }\textbf {\bibinfo {volume}
  {152}},\ \bibinfo {pages} {184108} (\bibinfo {year} {2020})}\BibitemShut
  {NoStop}%
\bibitem [{\citenamefont {Kottmann}\ \emph
  {et~al.}(2020{\natexlab{c}})\citenamefont {Kottmann}, \citenamefont
  {Bischoff},\ and\ \citenamefont {Valeev}}]{kottmann2020direct}%
  \BibitemOpen
  \bibfield  {author} {\bibinfo {author} {\bibfnamefont {J.~S.}\ \bibnamefont
  {Kottmann}}, \bibinfo {author} {\bibfnamefont {F.~A.}\ \bibnamefont
  {Bischoff}},\ and\ \bibinfo {author} {\bibfnamefont {E.~F.}\ \bibnamefont
  {Valeev}},\ }\bibfield  {title} {\bibinfo {title} {Direct determination of
  optimal pair-natural orbitals in a real-space representation: The
  second-order {Moller--Plesset} energy},\ }\href
  {https://doi.org/10.1063/1.5141880} {\bibfield  {journal} {\bibinfo
  {journal} {J. Chem. Phys.}\ }\textbf {\bibinfo {volume} {152}},\ \bibinfo
  {pages} {074105} (\bibinfo {year} {2020}{\natexlab{c}})}\BibitemShut
  {NoStop}%
\bibitem [{\citenamefont {Chien}\ and\ \citenamefont
  {Whitfield}(2020)}]{chien2020custom}%
  \BibitemOpen
  \bibfield  {author} {\bibinfo {author} {\bibfnamefont {R.~W.}\ \bibnamefont
  {Chien}}\ and\ \bibinfo {author} {\bibfnamefont {J.~D.}\ \bibnamefont
  {Whitfield}},\ }\href@noop {} {\bibinfo {title} {Custom fermionic codes for
  quantum simulation}} (\bibinfo {year} {2020}),\ \Eprint
  {https://arxiv.org/abs/2009.11860} {arXiv:2009.11860 [quant-ph]} \BibitemShut
  {NoStop}%
\bibitem [{\citenamefont {Derby}\ and\ \citenamefont
  {Klassen}(2020)}]{derby2020compact}%
  \BibitemOpen
  \bibfield  {author} {\bibinfo {author} {\bibfnamefont {C.}~\bibnamefont
  {Derby}}\ and\ \bibinfo {author} {\bibfnamefont {J.}~\bibnamefont
  {Klassen}},\ }\href@noop {} {\bibinfo {title} {A compact fermion to qubit
  mapping}} (\bibinfo {year} {2020}),\ \Eprint
  {https://arxiv.org/abs/2003.06939} {arXiv:2003.06939 [quant-ph]} \BibitemShut
  {NoStop}%
\bibitem [{\citenamefont {Bravyi}\ \emph {et~al.}(2017)\citenamefont {Bravyi},
  \citenamefont {Gambetta}, \citenamefont {Mezzacapo},\ and\ \citenamefont
  {Temme}}]{bravyi2017tapering}%
  \BibitemOpen
  \bibfield  {author} {\bibinfo {author} {\bibfnamefont {S.}~\bibnamefont
  {Bravyi}}, \bibinfo {author} {\bibfnamefont {J.~M.}\ \bibnamefont
  {Gambetta}}, \bibinfo {author} {\bibfnamefont {A.}~\bibnamefont
  {Mezzacapo}},\ and\ \bibinfo {author} {\bibfnamefont {K.}~\bibnamefont
  {Temme}},\ }\bibfield  {title} {\bibinfo {title} {Tapering off qubits to
  simulate fermionic hamiltonians},\ }\href {https://arxiv.org/abs/1701.08213}
  {\bibfield  {journal} {\bibinfo  {journal} {arXiv:1701.08213 [quant-ph]}\ }
  (\bibinfo {year} {2017})}\BibitemShut {NoStop}%
\bibitem [{\citenamefont {Romero}\ \emph {et~al.}(2018)\citenamefont {Romero},
  \citenamefont {Babbush}, \citenamefont {McClean}, \citenamefont {Hempel},
  \citenamefont {Love},\ and\ \citenamefont
  {Aspuru-Guzik}}]{romero2018strategies}%
  \BibitemOpen
  \bibfield  {author} {\bibinfo {author} {\bibfnamefont {J.}~\bibnamefont
  {Romero}}, \bibinfo {author} {\bibfnamefont {R.}~\bibnamefont {Babbush}},
  \bibinfo {author} {\bibfnamefont {J.~R.}\ \bibnamefont {McClean}}, \bibinfo
  {author} {\bibfnamefont {C.}~\bibnamefont {Hempel}}, \bibinfo {author}
  {\bibfnamefont {P.~J.}\ \bibnamefont {Love}},\ and\ \bibinfo {author}
  {\bibfnamefont {A.}~\bibnamefont {Aspuru-Guzik}},\ }\bibfield  {title}
  {\bibinfo {title} {Strategies for quantum computing molecular energies using
  the unitary coupled cluster ansatz},\ }\href
  {https://doi.org/10.1088/2058-9565/aad3e4} {\bibfield  {journal} {\bibinfo
  {journal} {Quantum Sci. Technol.}\ }\textbf {\bibinfo {volume} {4}},\
  \bibinfo {pages} {014008} (\bibinfo {year} {2018})}\BibitemShut {NoStop}%
\bibitem [{\citenamefont {Gard}\ \emph {et~al.}(2020)\citenamefont {Gard},
  \citenamefont {Zhu}, \citenamefont {Barron}, \citenamefont {Mayhall},
  \citenamefont {Economou},\ and\ \citenamefont {Barnes}}]{gard2020efficient}%
  \BibitemOpen
  \bibfield  {author} {\bibinfo {author} {\bibfnamefont {B.~T.}\ \bibnamefont
  {Gard}}, \bibinfo {author} {\bibfnamefont {L.}~\bibnamefont {Zhu}}, \bibinfo
  {author} {\bibfnamefont {G.~S.}\ \bibnamefont {Barron}}, \bibinfo {author}
  {\bibfnamefont {N.~J.}\ \bibnamefont {Mayhall}}, \bibinfo {author}
  {\bibfnamefont {S.~E.}\ \bibnamefont {Economou}},\ and\ \bibinfo {author}
  {\bibfnamefont {E.}~\bibnamefont {Barnes}},\ }\bibfield  {title} {\bibinfo
  {title} {Efficient symmetry-preserving state preparation circuits for the
  variational quantum eigensolver algorithm},\ }\href
  {https://doi.org/10.1038/s41534-019-0240-1} {\bibfield  {journal} {\bibinfo
  {journal} {NPJ Quantum Inf.}\ }\textbf {\bibinfo {volume} {6}},\ \bibinfo
  {pages} {1} (\bibinfo {year} {2020})}\BibitemShut {NoStop}%
\bibitem [{\citenamefont {Yalouz}\ \emph {et~al.}(2021)\citenamefont {Yalouz},
  \citenamefont {Senjean}, \citenamefont {Günther}, \citenamefont {Buda},
  \citenamefont {O'Brien},\ and\ \citenamefont
  {Visscher}}]{yalouz2020stateaveraged}%
  \BibitemOpen
  \bibfield  {author} {\bibinfo {author} {\bibfnamefont {S.}~\bibnamefont
  {Yalouz}}, \bibinfo {author} {\bibfnamefont {B.}~\bibnamefont {Senjean}},
  \bibinfo {author} {\bibfnamefont {J.}~\bibnamefont {Günther}}, \bibinfo
  {author} {\bibfnamefont {F.}~\bibnamefont {Buda}}, \bibinfo {author}
  {\bibfnamefont {T.~E.}\ \bibnamefont {O'Brien}},\ and\ \bibinfo {author}
  {\bibfnamefont {L.}~\bibnamefont {Visscher}},\ }\bibfield  {title} {\bibinfo
  {title} {A state-averaged orbital-optimized hybrid
  quantum{\textendash}classical algorithm for a democratic description of
  ground and excited states},\ }\href
  {https://doi.org/10.1088/2058-9565/abd334} {\bibfield  {journal} {\bibinfo
  {journal} {Quantum Sci. Technol.}\ }\textbf {\bibinfo {volume} {6}},\
  \bibinfo {pages} {024004} (\bibinfo {year} {2021})}\BibitemShut {NoStop}%
\bibitem [{\citenamefont {Sokolov}\ \emph {et~al.}(2020)\citenamefont
  {Sokolov}, \citenamefont {Barkoutsos}, \citenamefont {Ollitrault},
  \citenamefont {Greenberg}, \citenamefont {Rice}, \citenamefont {Pistoia},\
  and\ \citenamefont {Tavernelli}}]{sokolov2020}%
  \BibitemOpen
  \bibfield  {author} {\bibinfo {author} {\bibfnamefont {I.~O.}\ \bibnamefont
  {Sokolov}}, \bibinfo {author} {\bibfnamefont {P.~K.}\ \bibnamefont
  {Barkoutsos}}, \bibinfo {author} {\bibfnamefont {P.~J.}\ \bibnamefont
  {Ollitrault}}, \bibinfo {author} {\bibfnamefont {D.}~\bibnamefont
  {Greenberg}}, \bibinfo {author} {\bibfnamefont {J.}~\bibnamefont {Rice}},
  \bibinfo {author} {\bibfnamefont {M.}~\bibnamefont {Pistoia}},\ and\ \bibinfo
  {author} {\bibfnamefont {I.}~\bibnamefont {Tavernelli}},\ }\bibfield  {title}
  {\bibinfo {title} {Quantum orbital-optimized unitary coupled cluster methods
  in the strongly correlated regime: Can quantum algorithms outperform their
  classical equivalents?},\ }\href {https://doi.org/10.1063/1.5141835}
  {\bibfield  {journal} {\bibinfo  {journal} {J. Chem. Phys.}\ }\textbf
  {\bibinfo {volume} {152}},\ \bibinfo {pages} {124107} (\bibinfo {year}
  {2020})}\BibitemShut {NoStop}%
\bibitem [{\citenamefont {Lee}\ \emph {et~al.}(2018)\citenamefont {Lee},
  \citenamefont {Huggins}, \citenamefont {Head-Gordon},\ and\ \citenamefont
  {Whaley}}]{lee2018generalized}%
  \BibitemOpen
  \bibfield  {author} {\bibinfo {author} {\bibfnamefont {J.}~\bibnamefont
  {Lee}}, \bibinfo {author} {\bibfnamefont {W.~J.}\ \bibnamefont {Huggins}},
  \bibinfo {author} {\bibfnamefont {M.}~\bibnamefont {Head-Gordon}},\ and\
  \bibinfo {author} {\bibfnamefont {K.~B.}\ \bibnamefont {Whaley}},\ }\bibfield
   {title} {\bibinfo {title} {Generalized unitary coupled cluster wave
  functions for quantum computation},\ }\href
  {https://doi.org/10.1021/acs.jctc.8b01004} {\bibfield  {journal} {\bibinfo
  {journal} {J. Chem. Theory Comput.}\ }\textbf {\bibinfo {volume} {15}},\
  \bibinfo {pages} {311} (\bibinfo {year} {2018})}\BibitemShut {NoStop}%
\bibitem [{\citenamefont {Ryabinkin}\ \emph {et~al.}(2018)\citenamefont
  {Ryabinkin}, \citenamefont {Yen}, \citenamefont {Genin},\ and\ \citenamefont
  {Izmaylov}}]{ryabinkin2018qubit}%
  \BibitemOpen
  \bibfield  {author} {\bibinfo {author} {\bibfnamefont {I.~G.}\ \bibnamefont
  {Ryabinkin}}, \bibinfo {author} {\bibfnamefont {T.-C.}\ \bibnamefont {Yen}},
  \bibinfo {author} {\bibfnamefont {S.~N.}\ \bibnamefont {Genin}},\ and\
  \bibinfo {author} {\bibfnamefont {A.~F.}\ \bibnamefont {Izmaylov}},\
  }\bibfield  {title} {\bibinfo {title} {Qubit coupled cluster method: a
  systematic approach to quantum chemistry on a quantum computer},\ }\href
  {https://doi.org/10.1021/acs.jctc.8b00932} {\bibfield  {journal} {\bibinfo
  {journal} {J. Chem. Theory Comput.}\ }\textbf {\bibinfo {volume} {14}},\
  \bibinfo {pages} {6317} (\bibinfo {year} {2018})}\BibitemShut {NoStop}%
\bibitem [{\citenamefont {Ryabinkin}\ \emph {et~al.}(2020)\citenamefont
  {Ryabinkin}, \citenamefont {Lang}, \citenamefont {Genin},\ and\ \citenamefont
  {Izmaylov}}]{ryabinkin2020iterative}%
  \BibitemOpen
  \bibfield  {author} {\bibinfo {author} {\bibfnamefont {I.~G.}\ \bibnamefont
  {Ryabinkin}}, \bibinfo {author} {\bibfnamefont {R.~A.}\ \bibnamefont {Lang}},
  \bibinfo {author} {\bibfnamefont {S.~N.}\ \bibnamefont {Genin}},\ and\
  \bibinfo {author} {\bibfnamefont {A.~F.}\ \bibnamefont {Izmaylov}},\
  }\bibfield  {title} {\bibinfo {title} {Iterative qubit coupled cluster
  approach with efficient screening of generators},\ }\href
  {https://doi.org/10.1021/acs.jctc.9b01084} {\bibfield  {journal} {\bibinfo
  {journal} {J. Chem. Theory Comput.}\ }\textbf {\bibinfo {volume} {16}},\
  \bibinfo {pages} {1055} (\bibinfo {year} {2020})}\BibitemShut {NoStop}%
\bibitem [{\citenamefont {Lang}\ \emph {et~al.}(2020)\citenamefont {Lang},
  \citenamefont {Ryabinkin},\ and\ \citenamefont
  {Izmaylov}}]{lang2020iterative}%
  \BibitemOpen
  \bibfield  {author} {\bibinfo {author} {\bibfnamefont {R.~A.}\ \bibnamefont
  {Lang}}, \bibinfo {author} {\bibfnamefont {I.~G.}\ \bibnamefont
  {Ryabinkin}},\ and\ \bibinfo {author} {\bibfnamefont {A.~F.}\ \bibnamefont
  {Izmaylov}},\ }\bibfield  {title} {\bibinfo {title} {Unitary transformation
  of the electronic hamiltonian with an exact quadratic truncation of the
  baker-campbell-hausdorff expansion},\ }\href
  {https://arxiv.org/abs/2002.05701v2} {\bibfield  {journal} {\bibinfo
  {journal} {arXiv:2002.05701 [quant-ph]}\ } (\bibinfo {year}
  {2020})}\BibitemShut {NoStop}%
\bibitem [{\citenamefont {Grimsley}\ \emph
  {et~al.}(2019{\natexlab{a}})\citenamefont {Grimsley}, \citenamefont
  {Economou}, \citenamefont {Barnes},\ and\ \citenamefont
  {Mayhall}}]{grimsley2019adaptive}%
  \BibitemOpen
  \bibfield  {author} {\bibinfo {author} {\bibfnamefont {H.~R.}\ \bibnamefont
  {Grimsley}}, \bibinfo {author} {\bibfnamefont {S.~E.}\ \bibnamefont
  {Economou}}, \bibinfo {author} {\bibfnamefont {E.}~\bibnamefont {Barnes}},\
  and\ \bibinfo {author} {\bibfnamefont {N.~J.}\ \bibnamefont {Mayhall}},\
  }\bibfield  {title} {\bibinfo {title} {An adaptive variational algorithm for
  exact molecular simulations on a quantum computer},\ }\href
  {https://doi.org/10.1038/s41467-019-10988-2} {\bibfield  {journal} {\bibinfo
  {journal} {Nat. Commun.}\ }\textbf {\bibinfo {volume} {10}},\ \bibinfo
  {pages} {1} (\bibinfo {year} {2019}{\natexlab{a}})}\BibitemShut {NoStop}%
\bibitem [{\citenamefont {Tang}\ \emph {et~al.}(2019)\citenamefont {Tang},
  \citenamefont {Barnes}, \citenamefont {Grimsley}, \citenamefont {Mayhall},\
  and\ \citenamefont {Economou}}]{tang2019qubit}%
  \BibitemOpen
  \bibfield  {author} {\bibinfo {author} {\bibfnamefont {H.~L.}\ \bibnamefont
  {Tang}}, \bibinfo {author} {\bibfnamefont {E.}~\bibnamefont {Barnes}},
  \bibinfo {author} {\bibfnamefont {H.~R.}\ \bibnamefont {Grimsley}}, \bibinfo
  {author} {\bibfnamefont {N.~J.}\ \bibnamefont {Mayhall}},\ and\ \bibinfo
  {author} {\bibfnamefont {S.~E.}\ \bibnamefont {Economou}},\ }\bibfield
  {title} {\bibinfo {title} {qubit-{ADAPT-VQE}: An adaptive algorithm for
  constructing hardware-efficient ansatze on a quantum processor},\ }\href
  {https://arxiv.org/abs/1911.10205} {\bibfield  {journal} {\bibinfo  {journal}
  {arXiv:1911.10205 [quant-ph]}\ } (\bibinfo {year} {2019})}\BibitemShut
  {NoStop}%
\bibitem [{\citenamefont {Kottmann}\ \emph
  {et~al.}(2021{\natexlab{b}})\citenamefont {Kottmann}, \citenamefont {Anand},\
  and\ \citenamefont {Aspuru-Guzik}}]{kottmann2020feasible}%
  \BibitemOpen
  \bibfield  {author} {\bibinfo {author} {\bibfnamefont {J.~S.}\ \bibnamefont
  {Kottmann}}, \bibinfo {author} {\bibfnamefont {A.}~\bibnamefont {Anand}},\
  and\ \bibinfo {author} {\bibfnamefont {A.}~\bibnamefont {Aspuru-Guzik}},\
  }\bibfield  {title} {\bibinfo {title} {A feasible approach for automatically
  differentiable unitary coupled-cluster on quantum computers},\ }\href
  {https://doi.org/10.1039/D0SC06627C} {\bibfield  {journal} {\bibinfo
  {journal} {Chem. Sci.}\ ,\ } (\bibinfo {year}
  {2021}{\natexlab{b}})}\BibitemShut {NoStop}%
\bibitem [{\citenamefont {Grimsley}\ \emph
  {et~al.}(2019{\natexlab{b}})\citenamefont {Grimsley}, \citenamefont
  {Claudino}, \citenamefont {Economou}, \citenamefont {Barnes},\ and\
  \citenamefont {Mayhall}}]{grimsley2019trotterized}%
  \BibitemOpen
  \bibfield  {author} {\bibinfo {author} {\bibfnamefont {H.~R.}\ \bibnamefont
  {Grimsley}}, \bibinfo {author} {\bibfnamefont {D.}~\bibnamefont {Claudino}},
  \bibinfo {author} {\bibfnamefont {S.~E.}\ \bibnamefont {Economou}}, \bibinfo
  {author} {\bibfnamefont {E.}~\bibnamefont {Barnes}},\ and\ \bibinfo {author}
  {\bibfnamefont {N.~J.}\ \bibnamefont {Mayhall}},\ }\bibfield  {title}
  {\bibinfo {title} {Is the trotterized uccsd ansatz chemically
  well-defined?},\ }\href {https://doi.org/10.1021/acs.jctc.9b01083} {\bibfield
   {journal} {\bibinfo  {journal} {J. Chem. Theory Comput.}\ } (\bibinfo {year}
  {2019}{\natexlab{b}})}\BibitemShut {NoStop}%
\bibitem [{\citenamefont {Izmaylov}\ \emph {et~al.}(2020)\citenamefont
  {Izmaylov}, \citenamefont {Díaz-Tinoco},\ and\ \citenamefont
  {Lang}}]{Izmaylov2020ontheorder}%
  \BibitemOpen
  \bibfield  {author} {\bibinfo {author} {\bibfnamefont {A.~F.}\ \bibnamefont
  {Izmaylov}}, \bibinfo {author} {\bibfnamefont {M.}~\bibnamefont
  {Díaz-Tinoco}},\ and\ \bibinfo {author} {\bibfnamefont {R.~A.}\ \bibnamefont
  {Lang}},\ }\bibfield  {title} {\bibinfo {title} {On the order problem in
  construction of unitary operators for the variational quantum eigensolver},\
  }\href {https://doi.org/10.1039/D0CP01707H} {\bibfield  {journal} {\bibinfo
  {journal} {Phys. Chem. Chem. Phys.}\ }\textbf {\bibinfo {volume} {22}},\
  \bibinfo {pages} {12980} (\bibinfo {year} {2020})}\BibitemShut {NoStop}%
\bibitem [{\citenamefont {Higgott}\ \emph {et~al.}(2019)\citenamefont
  {Higgott}, \citenamefont {Wang},\ and\ \citenamefont
  {Brierley}}]{higgott2019variational}%
  \BibitemOpen
  \bibfield  {author} {\bibinfo {author} {\bibfnamefont {O.}~\bibnamefont
  {Higgott}}, \bibinfo {author} {\bibfnamefont {D.}~\bibnamefont {Wang}},\ and\
  \bibinfo {author} {\bibfnamefont {S.}~\bibnamefont {Brierley}},\ }\bibfield
  {title} {\bibinfo {title} {Variational quantum computation of excited
  states},\ }\href {https://doi.org/10.22331/q-2019-07-01-156} {\bibfield
  {journal} {\bibinfo  {journal} {Quantum}\ }\textbf {\bibinfo {volume} {3}},\
  \bibinfo {pages} {156} (\bibinfo {year} {2019})}\BibitemShut {NoStop}%
\bibitem [{\citenamefont {P{\'{e}}rez-Salinas}\ \emph
  {et~al.}(2020)\citenamefont {P{\'{e}}rez-Salinas}, \citenamefont
  {Cervera-Lierta}, \citenamefont {Gil-Fuster},\ and\ \citenamefont
  {Latorre}}]{QClass}%
  \BibitemOpen
  \bibfield  {author} {\bibinfo {author} {\bibfnamefont {A.}~\bibnamefont
  {P{\'{e}}rez-Salinas}}, \bibinfo {author} {\bibfnamefont {A.}~\bibnamefont
  {Cervera-Lierta}}, \bibinfo {author} {\bibfnamefont {E.}~\bibnamefont
  {Gil-Fuster}},\ and\ \bibinfo {author} {\bibfnamefont {J.~I.}\ \bibnamefont
  {Latorre}},\ }\bibfield  {title} {\bibinfo {title} {Data re-uploading for a
  universal quantum classifier},\ }\href
  {https://doi.org/10.22331/q-2020-02-06-226} {\bibfield  {journal} {\bibinfo
  {journal} {{Quantum}}\ }\textbf {\bibinfo {volume} {4}},\ \bibinfo {pages}
  {226} (\bibinfo {year} {2020})}\BibitemShut {NoStop}%
\bibitem [{\citenamefont {Ponce}\ \emph {et~al.}(2019)\citenamefont {Ponce},
  \citenamefont {van Zon}, \citenamefont {Northrup}, \citenamefont {Gruner},
  \citenamefont {Chen}, \citenamefont {Ertinaz}, \citenamefont {Fedoseev},
  \citenamefont {Groer}, \citenamefont {Mao}, \citenamefont {Mundim} \emph
  {et~al.}}]{niagara1}%
  \BibitemOpen
  \bibfield  {author} {\bibinfo {author} {\bibfnamefont {M.}~\bibnamefont
  {Ponce}}, \bibinfo {author} {\bibfnamefont {R.}~\bibnamefont {van Zon}},
  \bibinfo {author} {\bibfnamefont {S.}~\bibnamefont {Northrup}}, \bibinfo
  {author} {\bibfnamefont {D.}~\bibnamefont {Gruner}}, \bibinfo {author}
  {\bibfnamefont {J.}~\bibnamefont {Chen}}, \bibinfo {author} {\bibfnamefont
  {F.}~\bibnamefont {Ertinaz}}, \bibinfo {author} {\bibfnamefont
  {A.}~\bibnamefont {Fedoseev}}, \bibinfo {author} {\bibfnamefont
  {L.}~\bibnamefont {Groer}}, \bibinfo {author} {\bibfnamefont
  {F.}~\bibnamefont {Mao}}, \bibinfo {author} {\bibfnamefont {B.~C.}\
  \bibnamefont {Mundim}}, \emph {et~al.},\ }\bibfield  {title} {\bibinfo
  {title} {{Deploying a Top-100 Supercomputer for Large Parallel Workloads: the
  Niagara Supercomputer}},\ }\href {https://doi.org/10.1145/3332186.3332195}
  {\bibfield  {journal} {\bibinfo  {journal} {Proceedings of the Practice and
  Experience in Advanced Research Computing on Rise of the Machines
  (learning)}\ ,\ \bibinfo {pages} {1}} (\bibinfo {year} {2019})}\BibitemShut
  {NoStop}%
\bibitem [{\citenamefont {Loken}\ \emph {et~al.}(2010)\citenamefont {Loken},
  \citenamefont {Gruner}, \citenamefont {Groer}, \citenamefont {Peltier},
  \citenamefont {Bunn}, \citenamefont {Craig}, \citenamefont {Henriques},
  \citenamefont {Dempsey}, \citenamefont {Yu}, \citenamefont {Chen} \emph
  {et~al.}}]{niagara2}%
  \BibitemOpen
  \bibfield  {author} {\bibinfo {author} {\bibfnamefont {C.}~\bibnamefont
  {Loken}}, \bibinfo {author} {\bibfnamefont {D.}~\bibnamefont {Gruner}},
  \bibinfo {author} {\bibfnamefont {L.}~\bibnamefont {Groer}}, \bibinfo
  {author} {\bibfnamefont {R.}~\bibnamefont {Peltier}}, \bibinfo {author}
  {\bibfnamefont {N.}~\bibnamefont {Bunn}}, \bibinfo {author} {\bibfnamefont
  {M.}~\bibnamefont {Craig}}, \bibinfo {author} {\bibfnamefont
  {T.}~\bibnamefont {Henriques}}, \bibinfo {author} {\bibfnamefont
  {J.}~\bibnamefont {Dempsey}}, \bibinfo {author} {\bibfnamefont {C.-H.}\
  \bibnamefont {Yu}}, \bibinfo {author} {\bibfnamefont {J.}~\bibnamefont
  {Chen}}, \emph {et~al.},\ }\bibfield  {title} {\bibinfo {title} {{SciNet:
  Lessons Learned from Building a Power-efficient Top-20 System and Data
  Centre}},\ }\href {https://doi.org/10.1088/1742-6596/256/1/012026} {\bibfield
   {journal} {\bibinfo  {journal} {J. Phys. Conf. Ser.}\ }\textbf {\bibinfo
  {volume} {256}},\ \bibinfo {pages} {012026} (\bibinfo {year}
  {2010})}\BibitemShut {NoStop}%
\end{thebibliography}%
\clearpage
\end{document}